%% file: CRB_gains.tex
\documentclass[fleqn,usenatbib,useAMS]{mnras}
\usepackage{hyperref}
\usepackage{amsmath}
\usepackage{array}
\usepackage{graphicx}
\usepackage{epstopdf}
\usepackage{enumerate}
\usepackage{subfig}
\usepackage{float}

\include{notations}

\newcommand{\Jones}[0]{\bGamma}

\title[On the calibratability of radio interferometers]{On Identifiability and Estimability of Direction Dependent Calibration of Radio Interferometric Arrays}

\author[A. Mouri Sardarabadi \& L.V.E. Koopmans]{
	Ahmad Mouri Sardarabadi\thanks{E-mail: ammsa@astro.rug.nl}
	and L\'eon V.E. Koopmans
	\\~\\
	Kapteyn Astronomical Institute, University of Groningen, P.O.Box 800, 9700 AV Groningen, The Netherlands.\\
}

\pubyear{2017}

\begin{document}
	
	\maketitle
	\begin{abstract} Calibration is a key step in the signal processing pipeline of any radio astronomical instrument. The required sky, ionospheric and instrumental models for this step can suffer from various kinds of incompleteness. In this paper we analyze several important calibration methods, ignoring for now the ionosphere. The aim is to use established statistical and signal processing tools to provide a generic method to assess calibratability of an instrument. We show how currently popular calibration techniques differ in their assumptions and also discuss their theoretical commonalities. We also study the effect of only using a sub-set of baselines on the calibration and provide theoretical methods to analyze excess noise and biases that it might introduce. In order to simplify the physical interpretation of the results, we introduce a new signal processing model which is capable of modeling instrumental direction dependent effects and spectral smoothness of the individual receiver gain within a beam-formed station. The statistical properties of this model are then studied by deriving the Cram\'er--Rao bound (CRB). We finally define a mathematical framework for calibratability of an instrument based on the model used which is generic and can be used to study different instruments. These theoretical results are then verified using numerical simulations. 
	\end{abstract}
	
	
	%%%%%%%%%%%%%%%%%%%%%%%%%%%%%%%%%%%%%%%%%%%%%%%%%%%%%%%%%%%%%%%%%%%
	
	\section{Introduction}
	
	One of the most challenging scientific drivers of current and future low-frequency radio telescopes such as LOFAR \citep{Haarlem2013}, SKA \citep{Dewdney2009,Hall2005,Koopmans2015} and HERA \citep{DeBoer2017} is the study of the 21-cm signal of neutral hydrogen from the Epoch of Reionization (EoR) and the Cosmic Dawn \citep{Furlanetto2006a,Morales2010}. The feeble 21-cm signal from these Cosmic eras requires these telescopes to achieve an extremely high dynamic range, ($>$$10^6$) between the signal and the brightest compact sources, over a wide field of view. 
	%
	%Another challenge is the occupation of the spectrum with other radiators which must be removed. 
	In order to remove the brightest compact sources, an accurate model of their positions and fluxes has to be obtained, typically down to mJy brightness levels. We call this the `sky model' hereafter. This can be done using various imaging techniques \citep{Hoegbom1974, Offringa2014, Carrillo2014, Leshem2000a, Leshem2009, MouriSardarabadi2015} . Once such a sky model exists, it can also be used to improve the calibration parameters of the instrument, in particular the complex receiver gains, which can be different per receiver and frequency channel \citep{Boonstra2003, Kazemi2013, Wijnholds2009, MouriSardarabadi2014}.
	With such improved calibration of the instrument, subsequently more sources might be detected after imaging, including the fainter diffuse foreground emission (mostly coming from the Milky Way). Hence during each calibration step, the sky model is still incomplete and it also contain errors due to calibration. The combination of the two steps can be seen as an alternating optimization technique which approaches monotonically to a (global) solution. This method is known as `self-calibration' \citep{Tol2007}. 
	There are complications, however, in constructing a sky model. For example, the number of detectable sources rapidly increases for very small brightness levels. This places both practical and theoretical limitations on the number of sources that can be included in a sky model. The sources are also often not exactly point like, but can be slightly extended. There are various methods, such as wavelet/shapelets modeling or statistical modeling techniques, to address these unmodeled sources with reduced degree of freedom in order to include them during calibration \citep{Yatawatta2010}. Besides compact sources there is also very extended diffuse emission that is hard to model. Finally, many other choices need to be made such as: which baselines to include in the calibration step, the number of and directions in which to calibrate, the calibration solution time scale and the level of smoothness of the gain solutions in the frequency direction. In particular the latter is crucial because spectral smoothness of the sky and instrument is necessary in order to separate it for example from the faint 21-cm signal which varies spectrally \citep{MouriSardarabadi2018b}. 
	%
	%Sometimes other a-preiori information such as the spectral behavior of the certain type of unmodeled sources can be use to exclude them during calibration procedure \citep{mertens2017statistical}. 
	In this paper we are not trying to remedy these problems, since they will always exist to some level, but we will outline a signal processing framework that allows one to study their effects on the estimated quantities which are of interest to an astronomer and in particular study the theoretical bounds of these effect in order to test what level of accuracy can be achieved. To this end, we construct a general signal processing model in line with the work done in \citep{Wijnholds2009, Kazemi2013} and use this model for further statistical analyses. The new model includes frequency and direction dependent behavior of the synthesis aperture array, and allows for sky model incompleteness, choices about the number of direction to calibrate in, as well as choosing sub-sets of baselines used in calibration and imaging. We solely focus on gain calibration, being the main challenge in aperture array signal processing, and do not address the calibration of other parameters.
	While the original motivation for this research has been to study the effect of a baseline cut on direction dependent gain calibration in a hierarchical array, such as LOFAR, by precisely quantify the consequent theoretical increase of noise level in the residuals as observed by \citet{Patil2016}, we soon discovered that the observed increase is much higher than the level predicted by the cut alone and more general study of ``calibratability" was needed. 
	
	\noindent Throughout this manuscript we show how the study of uniqueness (or identifiability), estimability and bias is related to the study of calibratability. In Sec.~\ref{sec:calib_problem} we discuss the problem of gain calibration and show the commonalities and differences between different strategies currently popular. In Sec.~\ref{sec:models}, we show how these different assumptions translate into their equivalent signal processing models. In order to provide a better interpretation of the gains based of physics, we explicitly use the hierarchical structure of the telescope and model the direction dependent effects of the array as a function of gain variations on sub-station level, directly, e.g. HBA tiles in LOFAR \cite{Haarlem2013}. To our knowledge is the first time this model has been used to study calibratability in radio astronomy. We discuss the problem of uniqueness and identifiability for all of the introduced calibration models in Sec.~\ref{sec:identif}.  In Sec.~\ref{sec:CRB} and \ref{sec:incomp} we use the Cramer-Rao lower bound to estimate the theoretical minimum excess noise due to data exclusion and sky-model incompleteness and show its relation to the least square and Bayesian calibration. We show how all these different calibration methods can be approached in a unified way by using the concept of semi-linearity \citep{MouriSardarabadi2018b} in Sec.~\ref{sec:SL_EN_SP}. Finally, these theoretical results are used with reslasitic  simulations of the LOFAR and SKA telescope in Sec.~\ref{sec:SIM}.
	
	\section{Gain Calibration}
	\label{sec:calib_problem}
	
	In this section we provide an overview of various approaches to the calibration problem and the main assumptions they have. How these assumptions are translated into the mathematical models is discussed in the next section and throughout the paper.
	We define calibration as an estimation problem for the nuisance parameters in the data model (measurement equation). A nuisance parameter is an unknown which influences the measurements, but is not of interest for the final analysis of the data (e.g. the gain is a nuisance parameter, the desired 21-cm signal is not. However the former is needed to obtain the latter). This definition of calibration could be extended to include possible corrections for the undesired effects of these nuisance parameters.
	Using this definition, a key step in any calibration problem is the division of the parameter space into two sets. One set consisting of the desire parameters and a set of nuisance parameters which we call calibration parameters from this point forward. These sets can be very different depending on the science case under study and the assumptions on the instrument used to do the measurements.
	
	\subsection{Definitions}
	\label{subsec:def}
	
	In radio interferometry the instrument is an array of receivers. Each element of the array produces a voltage output which is a perturbed version of the desired (electromagnetic) signal. A common assumption in array processing is the narrow--band assumption, which allows for the modeling of geometrical delays between the receiving elements as phase changes \citep[see pp 23--24 in][]{MouriSardarabadi2016}. So the important measured quantities are the amplitudes and phases of the incoming signals on each receiver, and the product of amplitudes and the phase differences between different receivers. As a result, any process that changes these two quantities is considered a nuisance and part of the calibration problem. We divide these parameters into two classes: multiplicative and additive. For a single calibration problem the multiplicative changes to the signal are denoted as `gains' and additive perturbations are called `noise' if stochastic, and `bias' otherwise.
	As discussed above, the separation of the parameter space in two sets is a key step. In many radio-astronomical applications it is common to only define the calibrating parameters. The effect of these parameters is then mitigated and the resulting difference between the data and the model are called the `residuals'. The residuals are then assumed to be only a function of the desired parameters, and not the calibration parameters. If this turns out to be false, additional calibration steps are included. The underlying assumption is that the parameter space of the desired and calibrating signals are orthogonal. If this is not true, it leads to a signal `bias' which then requires additional calibration steps, or it leads to the suppression of the desired signal. 
	
	In this paper, we include the desired parameters (e.g.\ the sky signal) directly into the data model. This is discussed in Sect.\ \ref{sec:identif}. We also discuss suppression in Sect.\ \ref{sec:SL_EN_SP}.
	To further simply things, in this paper, we define the extended diffuse foreground emission and the much fainter 21-cm signal, which both are measured mainly using the shorter baselines, as our desired unknowns. The foreground removal done after this calibration step to separate the EoR signal from the extended foreground emission is not discussed here, and other techniques can be used for that \citep{Mertens2018}. We note that besides missing compact sources, these extended and diffuse desired signals are therefore also missing in the calibration model, and hence could lead to errors in the gain solutions. They could be included as a model as well, but the extended diffuse nature of the desired signal makes them hard to describe by a limited set of parameters, and hence often prohibitively expensive to calculate (see Sect.~\ref{subsec:limit}). 
	
	\subsection{Calibration Schemes}
	
	In this subsection we define the calibrating parameters under various different assumptions that are quite common in the literature. These schemes, and their implications, will be studied further in the remainder of this paper.
	Under ideal schemes we have the following assumptions,
	which we indicate as {\tt Scheme\,i} hereafter:
	
	\begin{enumerate}[(1)]
		
		\item {\sl Direction independent (DI) gains}: We assume that each receiver causes an independent amplitude and phase change to the signal which is exactly the same for all directions. Each receiver adds an independent Gaussian noise to the signal. We have a complete and accurate sky model that can be used to find an estimate for the gains and noise powers. This scheme was studied in e.g.\ \citet{Boonstra2003}.
		
		\item {\sl Direction independent gains with unknown source brightnesses:} We use the same assumptions as DI case with the modification that the sky model only includes the position of the sources but not their magnitude. This can also be regarded as a direction dependent gain calibration where all the receivers in the array have exactly the same directional response. An example can be found in  e.g.\ \citet{Wijnholds2009}.
		
		\item {\sl Effective direction dependent (DD) gains :} Each receiver (or beam-formed set of receivers) has an independent gain that can be different for certain number of directions. The gains do not necessarily have a physical interpretation and are a mixture of geometrical delays and other effects (e.g.\ ionosphere). A sky model in terms of the effective coherency matrix of each direction is assumed known. This was extensively studied in \citet{Kazemi2013}.
		
		\item {\sl Direction dependent gains with hierarchical beam-forming:} Each receiver is itself a beam-formed array. The direction dependent gains are modeled as weighted beam-forming of sub-receiver elements (see Sect.\ \ref{ssec:cov_beam_data}). A complete and accurate sky-model is available.
		
	\end{enumerate}
	In the current paper, we limit ourselves to these four schemes that cover much of the literature, but note that other more complex schemes are possible. For example, one can include the ionosphere as an additional direction, baseline, frequency, and spatially dependent gain effect \citep{Vedantham2016}. Currently, this effect can be `absorbed' in to the effective direction dependent gain solutions in {\tt Scheme\,3}, where beam and ionospheric effects are not distinguished. We defer the inclusion of a more physical ionospheric model to a future analysis.
	
	\subsection{Practical Limitations}\label{subsec:limit}
	
	\noindent In all of the above calibration schemes some or all of the following issues could exist:
	
	\begin{enumerate}[(a)]
		\item \textsl{Model incompleteness:} The sky model that is used for the calibration is not complete and has missing compact sources and/or diffuse emission.
		\item \textsl{Model inaccuracies:} The sky model is complete but is either obtained from noisy data itself and has errors in it components (e.g.\ wrong positions or fluxes).
		\item \textsl{Data incompleteness:} Part of the data is not available for calibration purposes (e.g.\ due to RFI, defective receivers, baseline cuts, etc.).
		\item \textsl{Model complexity:} The exact model cannot be used due to computational complexity constraints, and needs to be approximated by another effective model.
		\item \textsl{Calibratability:} The calibration problem is unidentifiable, which means that the desired parameters can not be uniquely estimated. One example is again the absorption of the desired parameters (e.g.\ diffuse foregrounds and 21-cm signal) in to the calibration parameters due to degeneracies (i.e.\ we call this `signal suppression').
	\end{enumerate}
	
	\noindent All these limitation, some of practical nature and others intrinsic to the problem at hand, will have some level of impact on the inferred desired parameters. To mitigate some of these effects, we can include additional regularization or constraints as priors on the gain solution, such as the smoothness of the gains as function frequency \citep{Brossard2018,MouriSardarabadi2018b,Yatawatta2016}. Having sketched the general schemes, assumptions and limitations, in the following sections we will now attach a precise mathematical signal/array processing model to each of these schemes and subsequently analyze them in further detail.
	
	\section{Signal Processing Models}
	\label{sec:models}
	In this section we introduce the various signal processing models, listed in the previous section, which are used in the analysis throughout this paper. We start with a common and simplified array processing model which ignores direction dependent effects, e.g.\ related to the beam and ionosphere. This model is based on the work presented by \citep{Boonstra2003, Wijnholds2009}. Many of the instrumentally related direction dependent gains (not those due to the ionosphere), however, are the result of adding multiple receiver signals (i.e.\ beam forming of an array of receivers in a station) each with their own direction independent gain error, before correlation. Using this, we show how to extend this direction independent model to include direction dependent effects as a result of beam-forming. moreover, if the station beam is much smaller than the receiver beam, we can ignore errors in the latter to first order, although some corrections will still be needed to account for the receiver beam (this receiver could in some cases be another beam-formed set of dipoles). For the notation used, we refer to Appendix~\ref{app:notation}.
	
	\subsection{Direction Independent Gains}
	\label{sec:SPM}
	
	With the above general context in mind, we assume to have access to the sampled voltage output of $P$ receivers (antennas, tiles of antennas, or stations, i.e.\ tiles of tiles). We stack these outputs in a $P \times 1$ vector denoted by $\bx$. We assume that the receivers are exposed to a set of point or compact sources, extended emission and noise. We also assume that the narrow--band assumption holds \citep[see pp 23--24 in][]{MouriSardarabadi2016}. This allows us to describe the output by
	\begin{equation}
	\label{eq:outputmodel}
	\bx(t)  = \sum_{q=1}^Q \bG(\bk_q)\ba(\bk_q) s_{ps,q}(t) + \bG_0\bs_r(t) +\bn(t)
	\end{equation}
	where $\bG(\bk)=\diag(\bg(\bk))$ models the complex receiver gains for directions $\bk$, $\ba(\bk)$ is a $P \times 1$ array response matrix for each source ($q=1 \dots Q$) and is a function of the geometric delays due to array topology, $s_{ps,q}(t)$ represents the signal from a point source (more complex compact source models such as shapelets are possible, modifying $\ba(\bk)$, as long as the direction dependent gains remain constant over the source), $\bG_0$ is a diagonal matrix modeling the direction independent gains of each receiver, $\bs_r(t)$ is the effective signal contribution of the extended sources to the array integrated over the entire sky including any direction dependent effects and $\bn$ is a $P \times 1$ vector modeling the signal contributions of the noise. Further we assume these signals to be zero mean with a Gaussian distribution. 
	In this section we assume that the instrument is only affected by direction independent gain perturbations, which means that $\bG(\bk)$ reduces to the diagonal matrix $\bG_0$ and is shared by all sources. Using this assumption and after taking the sampling of the output signal as function of time into account, we have
	\begin{equation}
	\bx[n]  = \bG (\bA \bs_{ps}[n] + \bs_r[n]) +\bn[n]
	\end{equation}
	where we dropped the subscript from $\bG_0$, $\bx[n]$ is the $n$th sample of the receivers output, $\bA$ is a $P \times Q$ matrix having  $\ba(\bk_q)$ as columns and $\bs_{ps}$ is a $Q \times 1$ vector formed by stacking the signals from the point sources. 
	We also assume the signals to be stationary during $N$ measurements where $N$ depends on the temporal and spectral resolution and stability of the instrument. This allows us to compute an estimate of the covariance matrix $\bR$ using these sampled data. This estimate is known as a sample covariance matrix or noisy visibility measurement and it is defined as
	\begin{equation}
	\bRh=\frac{1}{N}\sum_{n=1}^N \bx[n]\bx[n]^H.
	\end{equation}
	In radio interferometric imaging $\bx$ is rarely used directly and $\bRh$ is usually considered as the direct (correlator) output of the array. Because the direct output of the array, $\bx$, is to extremely accuracy a Gaussian signal \citep{Boonstra2003,Wijnholds2009}, the sample covariance matrix, $\bRh$, provides a sufficient statistic for analyzing the data. It must be emphasized here that removing the auto-correlations (diagonal part of $\bRh$), which is a common practice in radio interferometry, voids the statistical sufficiency and lads to loss of information. In this paper, we therefore retain the diagonal in the analysis. Using the sample covariance matrix as our noisy data, we model its expected value as follows
	\begin{align}
	\label{eq:DI_MODEL}
	\bR=\MCE\{\bRh\} &=\bG(\bA\bSigma_{\bs}\bA^H+\bR_r)\bG^H + \bR_{\bn} \notag\\
	& =\bG\bSigma\bG^H + \bG\bR_r\bG^H+ \bR_{\bn}
	\end{align}
	where  $\bSigma_{\bs}=\MCE\{\bs_{ps}\bs_{ps}^H \}=\diag(\bsigma_{ps})$ is the covariance matrix of the point sources and assumed to depend only on the intensity of the point sources $\bsigma_{ps}$, for which we assume to have a parametric model, $\bR_r=\MCE\{\bs_r\bs_r^H\}$ is the covariance matrix of any unmodeled (often the extended diffuse) emission, $\bR_{\bn}=\MCE\{\bn\bn^H\}$ is the noise covariance matrix and $\bSigma$ is the sky model. Without beamforming the sky model is given by \mbox{$\bSigma=\bA\bSigma_{\bs}\bA^H$} or equivalently, the covariance of the sky model without the influence of the gains (i.e.\ unit gains). This model can be extended to include more generic sky models such as compact sources, modeled by shapeless or wavelets, or even extended foreground emissions if a good model exists. We assume the noise contribution to be independent between the receivers and hence $\bR_{\bn}=\diag(\bsigma_{\bn})$ where $\bsigma_{\bn}$ is $P \times 1$ vector modeling the variance of the noise on each receiver. We also assume that the contribution of the extended emissions is limited to baselines shorter than $\alpha \lambda$ where $\alpha$ is a positive constant and $\lambda$ is the wavelength. If a source has significant contribution beyond this limit, it must be included in the sky model. One could set $\alpha=0$ if desired. Using a selection matrix, $\bS$, we can model the contribution of this unmodeled emission to the corresponding short baselines\footnote{Although we define the selection matrix here for shorter baselines that are are known to often experience diffuse emission that is unmodeled, one could define $\bS$ for any set of baselines where there might be unmodeled signal and in that way exclude those baselines in  the calibrating process.} as $\vect(\bR_r)=\bS\bsigma_r$. 
	Using this parametrization for shorter baselines, the model in vectorized form becomes
	\[
	\br = \vect(\bR) = (\bG^* \otimes \bG)[(\bA^* \circ \bA)\bsigma_{ps} + \bS\bsigma_r] + (\bI \circ \bI)\bsigma_{\bn},
	\]
	where $\otimes$ and $\circ$ represent the Kronecker and Khatri-Rao products respectively.
	Let $\bS$ and $\bD$ be a selection and a diagonal matrix  respectively, then there always exists a diagonal matrix $\bD_{\bS} = \bS^T\bD\bS$ such that $\bD \bS = \bS \bD_{\bS}$. Because the Kronecker product of two diagonal matrices is again diagonal, $\bG^* \otimes \bG$ is diagonal. Hence, without loss of generality, we can write
	\begin{equation}
	\label{eq:absorbed_gains_dic}
	\br = \vect(\bR) = (\bG^*\otimes \bG)\vect(\bSigma) + \bS\bG_{\bS}\bsigma_r + (\bI \circ \bI)\bsigma_{\bn},
	\end{equation}
	where \mbox{$\bG_{\bS} = \bS^T(\bG^*\otimes \bG)\bS$}.
	This relation shows that $\bsigma_r$ could absorb the direction independent effects of the unmodeled emission seen on short baselines (e.g.\ the diffuse foregrounds and 21-cm signal) into an effective $\bsigma_r' = \bG_{\bS}\bsigma_r$. Hence if we model the emission using a baseline cut, the direction independent gains can be estimated using the known sky model $\bSigma$ alone. We note that this calibration approach was used by \citet{Wijnholds2010a,MouriSardarabadi2014, Patil2016}, to avoid having to include bright diffuse emission into the sky model. Such an approach has considerable consequences, in particular leading to a discrete step in the visibility variance at the baseline length $\alpha \lambda$ \citep{Patil2016}. 
	
	The choice of the known and unknown parameters in the presented models is one of the key differences between the different calibration schemes presented in the previous section. As a result, it is important to explicitly discuss these differences.
	
	\subsubsection{Model Parameters}
	
	{\tt Scheme\,1}, introduced in the previous section, assumes that $\bSigma$ is completely known, in which case the calibrating parameters are the complex gains $\bg$, and the noise powers $\bsigma_{\bn}$ and the desired parameter is $\bsigma_{r}$. Hence
	\begin{equation}
	\label{eq:theta_DI}
	\btheta=\begin{bmatrix}
	\bg \\
	\bg^*\\
	\bsigma_{\bn}\\
	\bsigma_r
	\end{bmatrix}.
	\end{equation}
	In {\tt Scheme\,2}, on the other hand, $\bSigma$ is a function of the unknown source powers $\bsigma_{ps}$ which must be added in the calibrating parameters. This leads to the following vector of unknowns
	\begin{equation}
	\label{eq:theta_DI2}
	\btheta=\begin{bmatrix}
	\bg \\
	\bg^*\\
	\bsigma_{ps}\\
	\bsigma_{\bn}\\
	\bsigma_r
	\end{bmatrix}.
	\end{equation}
	We also introduce $\btheta_c$ and $\btheta_r$ as subsets of the vector $\btheta$ representing the calibrating and desired parameters, respectively. Since we are interested in the signals on the shorter baselines, e.g.\ the 21-cm signal, $\btheta_r = \bsigma_r$.  
	Because $\bx$ is Gaussian distributed we can find the lower bound for the covariance matrix of an unbiased estimator of $\btheta$ using the Cram\'er--Rao bound \citep[CRB; see pp 30--50 in][]{Kay1993a}. In sec.~\ref{sec:CRB}, we use this partitioning to find the CRB for each sub-set separately.
	
	\begin{figure}
		\centering
		\includegraphics[width=0.45\textwidth]{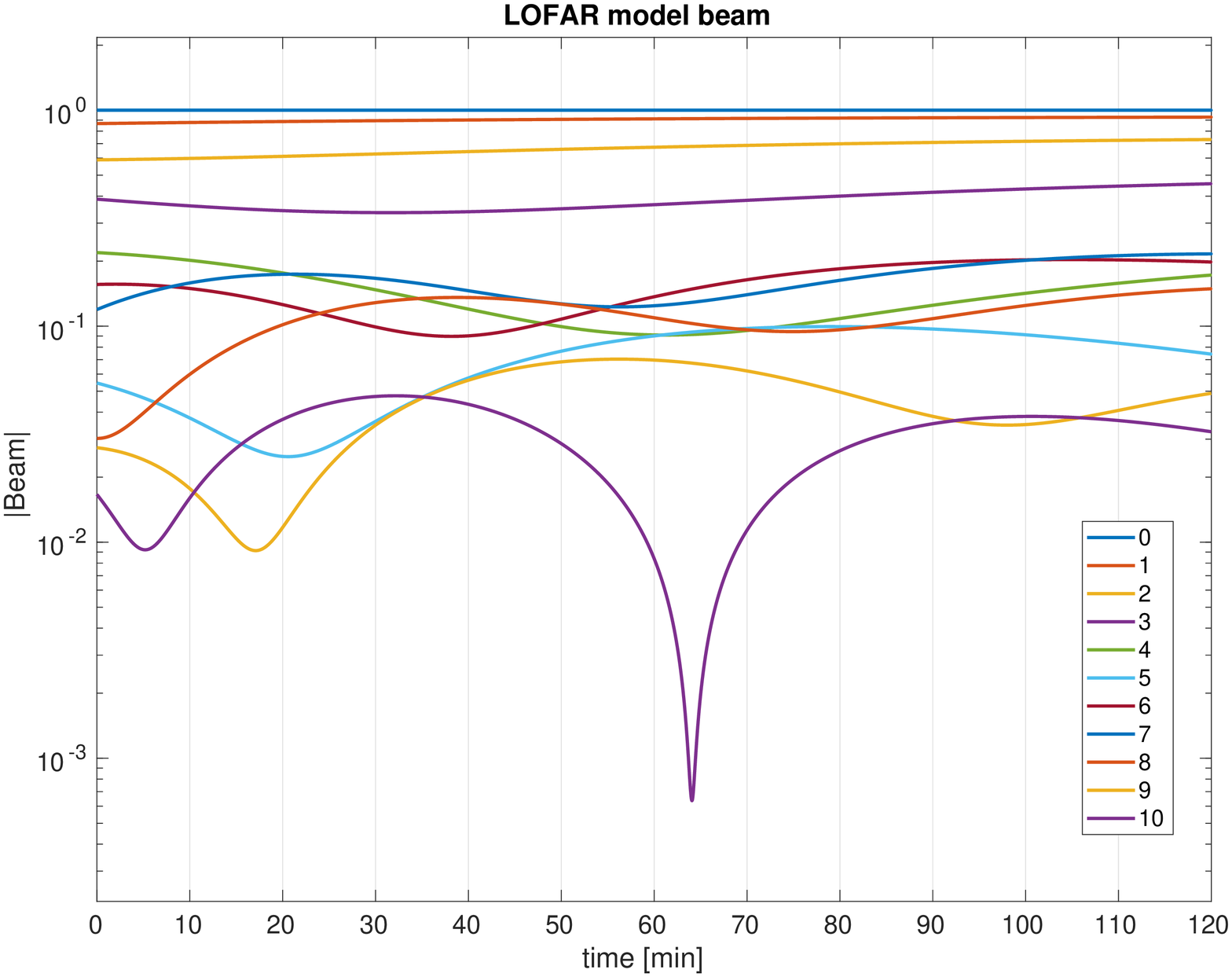}
		\caption{An example of the stability of the beam-formed gains $g_s$ as a function of time for a LOFAR High-Band Antenna station following the North Celestial Pole and a perfect receiver gain model ($\bG=\bI$). Eleven sources are simulated with distances between 0 and 5 degrees from the (phase) center of the field. Note that a source at 5 degrees from the phases center, passes through a null.}
		\label{fig:beam_stability}
	\end{figure}
	
	\subsection{Direction Dependent Gains}
	
	In this section we describe the various direction dependent gain models and their various approximations. In modern radio telescopes such as LOFAR, MWA and the future SKA, a single receiving element can be a beam-formed array of smaller receivers, creating a hierarchical system \citep{Haarlem2013,Hall2005}. Each receiver can be tracking a single or multiple fields. In this section we discuss the model for these systems, building on the result from the previous section.
	After applying a beam-former of the form $\bw(\bk_0)^H\bx$, where $\bw(\bk_0)$ is the beam-former for the field in the direction of $\bk_0$ and $\bx$ in \eqref{eq:outputmodel} we can show that an expression for the station gain towards a source at position $\bk$ can be written as
	\begin{align}
	\label{eq:staion_gain}
	g_s(\bk,\bk_0,\lambda)  &= \ba(\bk_0,\lambda)^H\bG(\bk,\lambda)\ba(\bk,\lambda)\notag \\
	&= \sum_p g_p(\bk,\lambda) e^{-\frac{2\text{j} \pi}{\lambda} \bzeta_p^T(\bk-\bk_0)}.
	\end{align}
	While this model is very useful for calculating the direction dependent gain of stations when $\bG(\bk,\lambda)$ is known, it is of little practical use if this gain function is unknown and needs to be estimated. Allowing the gains to attain independent values at each direction and frequency for each receiver leads to an ill-posed problem and increases the possibility of over-fitting the data \citep{vanderTol2009}. This means that $\bG(\bk,\lambda)$ needs to be further constrained or restricted. We have the option to put (physically motivated) constraints on a number of different ``dimensions'': on the direction of, or distance between, sources, on the time and/or frequency domain assuming there is some coherence over them, and on the dependency between individual dipoles (this could be interpreted as a spatial coherence). We discuss two possibilities and motivate the choice that is made for this work.
	
	\subsubsection{Simplified Models}
	
	The first choice could be a direct estimation of the gains at station level with no regard to the underlying elements. Physically one might regard this choosing a model for the direction dependent gains, and then interpreting its Fourier transform at the voltage patters of the station, regardless of whether this solution is physically plausible. In this scheme we effectively assume that all the elements in a station have the same direction dependent gain.
	\begin{align}
	\label{eq:staion_gain2}
	g_s(\bk,\bk_0,\lambda)  &= g(\bk,\lambda) \sum_p e^{-\frac{2\text{j} \pi}{\lambda} \bzeta_p^T(\bk-\bk_0)}\notag \\
	& =  g(\bk,\lambda)\ba(\bk_0,\lambda)^H\ba(\bk,\lambda).
	\end{align}
	One example of this direct approach is {\tt SageCal} (Scheme\,3 in Sect.\,\ref{sec:calib_problem}; \citet{Yatawatta2015}). One of the main advantages of this approach is its ability to model ionospheric effects together with beam errors, at the cost of reducing the ability to place physically motivated constraints on the solutions. In order to avoid degeneracy, some constrained could be included directly into the model. For example, we can assume that the gains for a sub-set of sources (i.e.\ for several $\bk$) are the same, because of their spacial proximity. We study this model and the corresponding identifiability problem in more details in Sec.~\ref{sec:identif}.
	%
	%A second choices is to allow all the dipoles within a station to have exactly the same direction dependent gains. Mathematically this translate into the following model $\bG(\bk,\lambda) = \bg(\bk,\lambda)\bI$ which translates to a station gain of the form
	%\[
	%g_s(\bk,\bk_0,\lambda) = \bg(\bk,\lambda) \ba(\bk_0,\lambda)^H\ba(\bk,\lambda).
	%\]
	%This is a common approximation where the station gain is defined as a product of the model station-beam, $\ba(\bk_0,\lambda)^H\ba(\bk,\lambda)$, and a direction dependent gain $\bg(\bk,\lambda)$.  One problem with this model is its limitations in capturing the change in the station-beam's shape as a result of the direction dependent gain variations. In order to clarify this statement we need to objectify the way we evaluate a ``good'' approximation of the beam-shape. One way to do this is by looking at the places where the beam becomes zero. This is important because it affects calibration in two ways. A wrong placement of a zero causes the calibrating algorithm to ignore sources which are not present and it boosts sources which are in the actual null of the beam. Because of the multiplicative nature of the model above the displacement of the beam nulls due to direction dependent effects can only be modeled in a limited way.
	%
	
	A second choice is to assume that the individual receivers have a direction independent gain. This is the assumptions under {\tt Scheme\,4} in Sect.\ \ref{sec:calib_problem}. Mathematically this translates to 
	\[
	g_s(\bk,\bk_0,\lambda) = \ba(\bk_0,\lambda)^H\bG(\lambda) \ba(\bk,\lambda).
	\]
	For stations where the resulting beam-formed station beam is much narrower than the beam of the individual (e.g.\ dipole) receivers this is often a fair assumption. Modeling the variations of the gains between different receivers allows for a better capturing of the beam-shape. However, smaller fluctuations and direction dependent effects which are not from the instrument (such as ionospheric effects) can not accurately be modeled using this model. 
	In both approaches, based on the number of stations, their layout/topology and other factors, the models could still remain degenerate and even further simplifications might be needed. For example, instead of calculating the gain of each receiver, a single gain can be calculated for a small number of receivers (e.g.\ a tile). This has the same effect as adding another layer of beam-forming similar to LOFAR High-Band Antennas \citep{Haarlem2013}. 
	Frequency dependency can also be added to both models which we discuss in Sect.\ \ref{ss:multichan}. We also assume that the beam is stable for a certain period of time. Fig.\ \ref{fig:beam_stability} shows the stability of the model beam for $\bG= \bI$ as a function of time for LOFAR HBA at $150$ MHz. This shows that for sources within the beam, the model beam is smooth and stable for a long period of time. For these sources the main reason for gain fluctuations (ignoring ionospheric effects) is the fluctuation in dipole gains. We assume the gain of each individual dipole is stable for typically several minutes. For sources near the nulls and further away from the center of the field the model beam changes much faster. If we include a beam model, this does not have any impact on the gain model. 
	The study done in this paper does not include any instrument independent fluctuations such as ionosphere which makes the second approximation sufficiently accurate for its purpose.
	
	\subsection{Covariance Model for DDC} 
	
	In this section we use the DD gain model introduced in \eqref{eq:staion_gain} to define two covariance models. The first model uses effective gains which are physically less motivated but could absorb ionospheric effects, the second being physically motivated but currently only able to solve for beam errors. 
	
	\subsubsection{Effective Direction-Dependent Gain Calibration}
	\label{ssec:SageCALModel}
	
	Below we discuss the signal model for {\tt Scheme\,3} using a single polarization (Stokes I), whereas the data model with polarization can be also treated as a direction dependent gain calibtration with twice the number of elements and directions \citep{sardarabadi2018efficient}. 
	This calibration technique, as implemented for LOFAR {\tt SageCal} \citep{Yatawatta2015}, can be seen as a very natural extension of the direction independent gain calibration. In this case, we divide the sky model into several source clusters and assume each cluster to have a single gain independent of direction inside the patch, this can be interpreted as performing classical self-calibration (e.g. \citet{Boonstra2003}) in multiple directions simultaneously. Hence, the data model is simply a sum over a set of direction independent solutions.
	Using a similar notation as in the previous sections we have the following covariance model \citep{Yatawatta2015,MouriSardarabadi2018b}
	\[ \bR = \sum_{q=1}^Q \bG_q \bSigma_q \bG_q^H + \bR_r + \bR_{\bn} \]
	where $q = 1,\dots,Q$ is the index for the cluster rather than each source in the sky model, $\bSigma_q$ is the sky model for all the sources in the cluster $q$ and $\bG_q$ is the gain for that cluster. We discuss the calibratability of this model in Sect.~\ref{sec:identif}.
	
	\subsubsection{Hierarchical Beam-formed Gain Calibration} \label{ssec:cov_beam_data}
	
	We derive a covariance model for the fourth calibration scheme. Assuming that a certain number of dipoles share the same gain. This is mathematically equivalent to a system where these dipoles are first beam-formed into single receiver before being beam-formed at the station level. Such a hierarchical system is found for example in LOFAR's High-Band Antenna configuration where the intermediate beam-formed receivers are called `tiles'. 
	%In this section we discuss a generic hierarchical system.
	%
	We assume that receiver $p$ at hierarchical level $l$ consists of $P_{l-1}$ elements. The output of each receiver is the result of a beam-forming done on its $P_{l-1}$ smaller receivers. We introduce the matrix $\bW_l$ for the $l$th level in the hierarchy such that the (sky) model for beam-formed system can be written as
	\[
	\bR_b = \bW_L \dots \bW_1\bA \bSigma_{\bs} \bA^H \bW_1^H \dots \bW_L^H
	\]
	where $\bR_b$ is the $P_L \times P_L$ beam-formed covariance matrix, and
	\[
	\bW_l=\begin{bmatrix}
	\bw^H_{l,1} &  \dots &  \zeros \\
	\vdots & \ddots  &   \zeros \\
	\zeros & \dots &  \bw^H_{l,P_l} \\
	\end{bmatrix}
	\]
	where, $\bw_{l,p}$ is a $P_{l-1} \times 1$ vector representing the beam-former for the field under observation. This hierarchy reduces the size of the array to that of $\bR_b$, hence considerably reducing the computational (and correlator) effort needed to model (and obtain) the data, at the price of loosing information over the larger field of view of a single receiver.  
	To simplify the notations, we introduce the effective beam-forming matrix \mbox{$\bW_b=\bW_L \dots \bW_2\bW_1$}. Using this notation we have 
	\[
	\begin{array}{rl}
	\bR &= \bW_b \bG(\bA\bSigma_{\bs}\bA^H + \bR_r + \bR_n)\bG^H \bW_b^H \\
	& = \bW_b \bG \bSigma \bG^H \bW_b^H+\bW_b\bG\bR_r\bG\bW_b^H  + \bW_b\bR_n \bW_b^H.
	\end{array}
	\]
	Note that if $P_{l-1}=1$ then $\bW_l=\bI$ hence we set $P_{-1}=1$ which is the lowest level in the hierarchy, i.e.\ each receiver consists of a single element and hence, $\bW_0=\bI$ always holds. 
	To illustrate this, we use LOFAR HBA as an example. A single station of LOFAR in HBA mode consists of $24$ or $48$ tiles. Each tile is an array of 16 receivers. In this case $L=2$ with $P_0=16$ and $P_1=24$ or $48$.
	We can swap the matrices $\bW_b$ and $\bG$ by using the following relations
	\begin{align}
	\bW_b\bG&=\begin{bmatrix}
	\bw^H_{1} &  \dots &  \zeros \\
	\vdots & \ddots  &   \zeros \\
	\zeros & \dots &  \bw^H_{P} \\
	\end{bmatrix}\begin{bmatrix}
	\bG_1 & \dots & \zeros \\
	\vdots & \ddots & \vdots \\
	\zeros & \dots & \bG_P
	\end{bmatrix} \notag\\
	&=\begin{bmatrix}
	\ones^T &  \dots &  \zeros \\
	\vdots & \ddots  &   \zeros \\
	\zeros & \dots &  \ones^T \\
	\end{bmatrix}\begin{bmatrix}
	\diag(\bw_1^*)\bG_1 & \dots & \zeros \\
	\vdots & \ddots & \vdots \\
	\zeros & \dots & \diag(\bw_P^*) \bG_P
	\end{bmatrix}\notag\\
	&=\begin{bmatrix}
	\ones^T &  \dots &  \zeros \\
	\vdots & \ddots  &   \zeros \\
	\zeros & \dots &  \ones^T \\
	\end{bmatrix}\bG\begin{bmatrix}
	\diag(\bw_1^*) & \dots & \zeros \\
	\vdots & \ddots & \vdots \\
	\zeros & \dots & \diag(\bw_P^*)
	\end{bmatrix}\notag\\
	&=\bB\bG\bWt_b
	\end{align}
	where \mbox{$\bWt_b = \diag([\bw_1^H, \dots, \bw_P^H ]^T)$} is now a diagonal matrix. Using this result we can again write the total beam-formed covariance matrix for a single snapshot as 
	\begin{equation}
	\label{eq:beamf_model}
	\bR= \bG_b \bSigma \bG_b^H + \bG_b \bWt_b \bR_r\bWt_b^H \bG_b^H  + \bW_b\bR_n\bW_b^H,
	\end{equation}
	where $\bG_b = \bB \bG$ and we define \mbox{$\bSigma \equiv \bWt_b \bA\bSigma_{\bs}\bA^H \bWt_b$} as the beamformed sky model. We show that the vectorized model for the short baselines keeps the same structure as the direction independent calibtation. This is done by using the properties of the selection matrix, which leads to
	\[
	\begin{array}{rl}
	\vect(\bG_b \bWt_b \bR_r\bWt_b^H \bG_b^H) &= (\bB \otimes \bB)(\bG^* \bWt_b^* \otimes \bG \bWt_b)\bS\bsigma_{r}\\
	&=(\bB \otimes \bB)\bS \bG_{\bS} \bsigma_{r} ,
	\end{array}
	\]
	with $\bS$ a selection matrix for the short baselines before the beamforming and \mbox{$\bG_{\bS} = \bS^H(\bG^* \bWt_b^* \otimes \bG \bWt_b)\bS$}  is a diagonal matrix, and its effects can be absorbed in $\bsigma_{r}$ similar to \eqref{eq:absorbed_gains_dic}.
	
	\noindent 
	Even though $\bG_b$ is constructed based on direction independent gain of each lower level receiver element in the hierarchy, it causes a direction dependent effect, unless the gains for all of the lower level elements are all equal. As a result, we define the average of the gains in a station as its `effective direction independent gain'. Based on the structure of the matrix $\bB$, we know that $\bD_{\bB} = \bB\bB^T$ is a $P \times P$ diagonal matrix where each of its diagonal elements is equal to the total number of beam-formed elements in the corresponding station. Hence, $\bG_0 = \bG_b\bB^T\bD_{\bB}^{-1}$ is a diagonal matrix representing the effective (average) direction independent gain of the stations.
	
	\subsubsection*{Relation between the Hirarchical and Effective Model}
	In the hirarchical model introduced in previous section, the gain matrix $\bG_b$ is no longer diagonal and therefore it has a direction dependent effect. As a result, from this point forward, the gain of a station is not characterized by a scalar, $g_p$, but by a vector, $\bgi_p$, or
	\[
	\bG_b = \begin{bmatrix}
	\bgi^H_{1} &  \dots &  \zeros \\
	\vdots & \ddots  &   \zeros \\
	\zeros & \dots &  \bgi^H_{P}
	\end{bmatrix}.
	\] 
	In order to show the flexibility of this model, we briefly discuss how it can be used to also produce the model used by {\tt Scheme\,3} in Sec.~\ref{ssec:SageCALModel}. For simplicity, let us assume that all stations consists of the same number of elements, which means that the size of $\bgi_p$ is the same for \mbox{$p=1,\dots,P$}. Let the length of $\bgi$ be denoted by $\hat{Q}$, in this case we have
	\[
	\bG_b\bK_{\hat{Q},P} = \begin{bmatrix}
	\bG_1,\dots,\bG_{\hat{Q}}
	\end{bmatrix},
	\]
	where $\bK_{N,M}$ is a permutation matrix such that $\vect(\bX^T)=\bK_{N,M} \vect(\bX)$ for any $N \times M$ matrix $\bX$ and $\bG_{\hat{q}}$ for \mbox{$\hat{q} = 1, \dots.\hat{Q}$} is a diagonal matrix obtained by collecting $\hat{q}$th element of each station into a diagonal matrix. We also define the permuted sky model as 
	\[
	\begin{array}{rl}
	\bSigmat &= \bK_{P,\hat{Q}} \bSigma \bK_{\hat{Q},P}\\
	& = \begin{bmatrix}
	\bSigmat_{1,1} & \dots & \bSigmat_{1,\hat{Q}}\\
	\vdots & \ddots & \vdots\\
	\bSigmat_{\hat{Q},1} & \dots & \bSigmat_{\hat{Q},\hat{Q}}
	\end{bmatrix}.
	\end{array}
	\]
	Using this permutation, we can rewrite \eqref{eq:beamf_model} as
	\[
	\bR = \sum_{i=1}^{\hat{Q}} \sum_{j=1}^{\hat{Q}} \bG_i \bSigmat_{i,j} \bG_j^H + \bG_b \bWt_b \bR_r\bWt_b^H \bG_b^H +\bW_b\bR_n\bW_b^H.
	\]
	Now, if we put $\hat{Q} = Q$ and $\bSigmat_{i,j} = \zeros$ for $i\neq j$, i.e. $\bSigmat_{i,i} = \bSigma_q$, we get the simplified model back. This also shows that calculating the beamformed model grows quadratic in the number of sub-elements, while the simplified model grows linearly with the number of directions. Also, because the sky model for the simplified model ignores all the cross terms, i.e. $\bSigmat_{i,j}$ with $i\neq j$, the gains are less constrained and can therefore absorb more gain fluctuations, such as ionosphere. The other way around, the use of the effective gains can be seen as creating $Q$ sub-arrays, containing a single element\footnote{The element can be "virtual" if the number of directions is more than the number of elements in each station.} from each station, which track different clusters. Clearly, the hierarchical model is a generalization of the simplified method and by choosing the right format for sky model, any theoretical performance result for this method is also valid for the other methods as well. As a result we only need to derive the CRB for the hierarchical model.
	\iffalse
	We re-emphasize  that in the case of effective gains model, the entire array factor (i.e.\ $\bA$ and $\bW_b$) was folded in to $\bSigma$ first and only then are effective gains applied per direction thereby mixing in the voltage pattern of the receiver array, in our beam-formed model on the other hand, the gains per single receiver (or tile) are first applied to the single-receiver level sky model and only then is the array factor applied. The latter operation thereby mixes the direction independent gains per receiver (having compact support in the station voltage pattern) in to direction dependent gains at the tile or station level. This is {\sl the} fundamental difference between the two approaches. However, the effective direction dependent gains are related to the receiver-based direction independent gains via a complex (Fourier-like) transformation and if many effective directions are used, it can lead to a localized gain correction of the size of a single receiver, thereby leading to very similar results. However, the approaches are quite distinct with a compact support gain/voltage pattern at the receiver level in the beam-formed model, and a compact-support gain pattern as function of direction (i.e.\ per finite patch of sky) in the effective direction dependent gain model.
	\fi
	
	\noindent
	Given the rapid growth in the number of unknowns, when multiple channels of data are processed, we need to take the frequency behavior of the gains into consideration and, if possible, use it to reduce the degeneracy in the model \citep{Yatawatta2016,MouriSardarabadi2018b,Brossard2018}. This is discussed in the next section.
	
	\subsection{Multi Frequency-Channel Model}
	\label{ss:multichan}
	
	As mentioned earlier, the gain model is often ill-posed due to the large number of unknown receiver or tile gains. To remedy this, we make use of the fact that the single receiver (i.e.\ amplifier) gains are in most cases extremely smooth as function of frequency \citep{Yatawatta2016,Brossard2018}, assuming the absence of cable reflections (although in principle this can be modeled in a similar manner). Hence, we can combine the covariance matrices of many frequency channels (up to hundreds in case of e.g.\ LOFAR, MWA and SKA) and adding only very few additional parameters to model the frequency behavior of the gains per receiver. Similarly, the foreground (model) is known to be spectrally smooth \citep{Mertens2018}, such that the number of parameters per model component also increases by much less than the number of frequency channels. Furthermore, given that baselines scale with wavelength -- densely sampling the $uvw$ space -- all this combined yields many more constraints on the gain model. Such a complete approach has for example been implemented in the consensus optimization extension of {\tt SageCal}, in the case of an effective gain model \citep{Yatawatta2015,Yatawatta2016}. Here we introduce this approach in our hierarchical model. 
	Here we use a similar method as \cite{MouriSardarabadi2018b} and assume to have access to sample covariance matrices for each frequency channel $k$ such that our dataset consists of $\bRh_k$ with $\bR_k=\MCE\{\bRh_k\}$ for $k=1,\dots,K$. By stacking all these snapshot into a single data vector we can model the entire dataset as
	\[
	\brh=\begin{bmatrix} \vect(\bRh_1) \\ \vdots \\ \vect(\bRh_K)\end{bmatrix}.
	\]
	As mentioned above, we assume that the gains and sky model are smooth functions of wavelength $\lambda$. One possible way to model the gains in this scheme is by introducing a `Vandermonde' matrix
	\[
	\bV_K^n=\begin{bmatrix}
	1 & \lambda_1 & \lambda_1^2& \dots & \lambda_1^n \\
	1 & \lambda_2 & \lambda_2^2& \dots & \lambda_2^n \\
	&  & \vdots&  &  \\
	1 & \lambda_K & \lambda_K^2& \dots & \lambda_K^n \\
	\end{bmatrix}.
	\]
	where $n$ is the order of the smooth polynomial model we are using for the gains, although any well-chosen (e.g.\ orthogonal basis) functional form could work. Using this matrix we can model the unknowns
	\begin{equation}
	\label{eq:mubetheta}
	\btheta=
	\begin{bmatrix}
	\bg_1^T &
	\dots &
	\bg_K^T&
	\bg_1^H &
	\dots &
	\bg_K^H&
	\bsigma^T_{r,1} &
	\dots &
	\bsigma^T_{r,K}
	\end{bmatrix}^T
	\end{equation}
	as
	\begin{equation}
	\label{eq:constrainedtheta}
	\btheta= \bVt \btheta_{\text{smooth}},
	\end{equation}
	where the vector $\btheta_{\text{smooth}}$ stacks the $n$ linear weights to the basis functions, and 
	\[
	\bVt = \left[ \begin{array}{c|c}
	\bI_2 \otimes \bV_K^n \otimes \bI_{P}  & \zeros \\ \hline
	\zeros & \bI
	\end{array}\right].
	\]
	
	\noindent
	Conversely, given any realization of $\btheta$ we can find the least-squares smooth version $\btheta_{\text{smooth}}$ using 
	\begin{equation}
	\label{eq:thetasmooth}
	\btheta_{\text{smooth}}=\left[ \begin{array}{c|c}
	\bI_2 \otimes [(\bV_K^n)^H\bV_K^n]^{-1}(\bV_K^n)^H \otimes \bI_{P}  & \zeros \\ \hline
	\zeros & \bI
	\end{array}\right ] \btheta.
	\end{equation}
	It is important to realize that \eqref{eq:thetasmooth} is exact and not an approximatin, because we assume \eqref{eq:constrainedtheta} to be true and exact for some value of $\btheta_{\rm smooth}$. In practice, the smoothness of the gains is much more important than the choice of the basis functions. The choice of polynomials is for simplicity and $\bVt$ can be replaced by any set of (sampled) basis functions and the general structure of model does not change \citep{MouriSardarabadi2018b}. Note that one advantage of using polynomials is that applying $[(\bV_K^n)^H\bV_K^n]^{-1}(\bV_K^n)^H$ to a vector can be done fast using interpolation techniques. 
	One problem that we might be confronted with is numerical stability for large wavelengths and higher order polynomial fitting (i.e. when $n$ is relatively large). Considering that the objective is the polynomial smoothness of the fit, without loss of generality, we can normalize the wavelengths to be in the interval $[-1,1]$. 
	Another issue that needs some attention follows from the assumption that the gains for each receiver are independent i.e. the coefficients for the polynomials in $\btheta_{\rm smooth}$ are modeled as random variables. The gains generated using this procedure have a much higher variation on the two boundaries of the frequency range than gains inside. 

	\section{Analysis of the Gain Models}
	
	Having completed our model descriptions of various popular gain models, and introduced a new spectrally-smooth  hierarchical beam-formed gain model, we are now in a situation to start analyzing these models in greater detail. Of particular interest is whether the model is at all calibratable. In other words, there are sufficient constraints to solve for all unknowns. This is of particular interest for arrays such as LOFAR and SKA that (will) solve for gains in many directions or for many receivers, and for many stations. For HERA this might not have to be done because the beam is formed by nearly identical dishes \citep{DeBoer2017}, but it might suffer from  a limited sky model. We will discuss this problem in the next subsections. 
	
	\subsection{Identifiability: How many gains can be calibrated per solution interval?}
	\label{sec:identif}
	
	We call a problem (locally) identifiable if the parameters of $\bR$ are uniquely determined by the data. If the problem is not identifiable, addition constraints must be added if a unique solution is desirable (or the chosen algorithm must have the inherent property which enforces such constraints). In some situations the quality of the solution does not depend on the chosen constraints, however choosing one is necessary. We make this more clear by addressing the identifiability issues for each of the problems above. Identifiability is a necessary condition for calibratability. However, it is not a sufficient condition and other factors such as signal-to-noise must also be considered when calibratability is discussed.
	While for simple models the identifiability issues could be spotted directly, in majority of the models it is not easy to see that the model suffers from an identifiability problem directly and additional mathematical tools are needed. It can be shown \citep{Rothenberg1971} that for data generated from any exponential family of probability distribution functions, the identifiability can be studied (locally) by investigating the singularity of the Fisher information matrix. In case of a Gaussian distribution this study can be further simplified by using the Jacobian of the measurement equation \citep{Schreier2010} instead. The Jacobian in this case is defined as
	\[
	\bJ = \frac{\partial \br}{\partial \btheta^T},
	\]
	where $\br$ is a vector obtained by stacking all of the (model) visibilities and $\btheta$ is a vector obtained by all of the unknowns. The rank deficiency of the Jacobian then tells us how many constraints are needed. We will apply this below.
	
	\subsubsection{Direction independent gains}
	
	For the direction independent gain calibration with known sky-model ({\tt Scheme\,1}), we know that if $\bg$ is a solution so is $\bg' = e^{\text{j}\bphi} \bg$ for any $\bphi$. This is the well-known phase ambiguity of direction independent gain solutions. If we also estimate the brightnesses of the sources, then additional ambiguity is introduced between the flux scale and the average gain. Let $\bg$ be a solution then $\sqrt{\balpha}\bg$ is also a solution because a factor $1/\balpha$ can be absorbed in the source brightnesses. Even if finding a solution for these identifiability problems is trivial, and many possibilities might exist, choosing one is necessary.
	%e
	While the identifiability of this calibration problem is simple enough to detect, for completeness we show how the Jacobian can be used to come to the same conclusion. We show that for direction independent gain calibration with a known sky model, the Jacobian is at least rank deficient by one, which as we know is the extra phase constraint that is needed. Using the definition of the Jacobian, \eqref{eq:DI_MODEL} and \eqref{eq:theta_DI} we have
	\[
	\bJ = \begin{bmatrix}
	\bG^H \bSigma^T \circ \bI_{P} & \bI_{P} \circ \bG\bSigma & \bS
	\end{bmatrix}
	\]
	where $\bSigma$ is the sky model. We can show that this matrix is at least rank deficient by 1 if we find a vector such that $\bJ\bz = \zeros$ and $\bz \neq \zeros$. Let
	\[
	\bz = \begin{bmatrix}
	\bg\\ -\bg^* \\ \zeros
	\end{bmatrix}
	\]
	then 
	\begin{equation}
	\label{eq:null_space_J_DI}
	\bJ\bz = \vect(\bG\bSigma\bG^H - \bG\bSigma\bG^H) = \zeros,
	\end{equation}
	for all $\bg \neq \zeros$. Hence $\bJ$ is always rank deficient by at least 1 which shows the claim that at least 1 constraint is necessary. In scheme 2 for which the unknowns are given by \eqref{eq:theta_DI} we have
	\[
	\bJ = \begin{bmatrix}
	\bG^H \bSigma^T \circ \bI_{P} & \bI_{P} \circ \bG\bSigma& \bG^H\bA^* \circ \bG\bA & \bS
	\end{bmatrix}
	\]
	and 
	\[
	\bZ = \begin{bmatrix}
	\bg & \bg\\ -\bg^* & \bg^* \\ \zeros & -\bsigma_{ps} \\\zeros & \zeros
	\end{bmatrix}.
	\]
	In this case $\bZ$ is a matrix with two (orthogonal) columns, showing that the Jacobian is at least rank deficient by 2. Again demonstrating that at least two constraints are needed to make the problem identifiable.
	If (similar to these examples) the exact choice of the constraints is not important, then algorithms could be developed based on the Jacobian that have constraint enforcing properties. Discussing such algorithms is beyond the scope of this paper. It is trivial to verify that for direction independent gain model above, as long as there are no defective receivers, i.e. $g_p \neq 0$ for $p = 1,\dots,P$, the rank of the Jacobian is completely defined by the sky model $\bSigma$ and the baseline cut which is modeled by $\bS$.
	For direction dependent problems, these identifiability issues could be more challenging to solve. We take a look at the gain solutions that use effective gains first and come back to the proposed method later in this paper.
	
	\subsubsection{Direction dependent effective gains}
	
	In order to analyze the identifiability of the effective gains, we divide the source clusters (simply called `clusters' hereafter) into two groups, a group of clusters with only a single (point) source in each of them and the rest of the clusters. Let $Q_1$ be the number clusters in the first group and $Q_2 = Q - Q_1$ be the number of clusters in the second group. It can be shown that the combining the clusters in the first group would lead to a Factor Analysis problem and requires $Q^2$ constraints \citep{MouriSardarabadi2016} for identifiability. The rest of the clusters are similar to the direction independent calibration requiring $Q_2$ phases to be fixed. In order to give a bound for the maximum number of clusters that could be estimated we make the simplification that all of the clusters have more than one source. In this case we have a necessary condition for identifiability that the degree of freedom, denoted by $s$, to be larger than zero. For single channel estimation without regularization we have $P^2-P$ known measurement points ($P$ auto-correlations have been removed), $2PQ$ unknown parameters for the gains and $Q$ constraints which leads to $s = P^2 - P - 2PQ + Q > 0$. Solving for $Q$ we have $Q < P(P-1)/(2P - 1) \approx (P-1)/2$. Hence,
	
	\begin{itemize}
		\item \sl For single frequency channel calibration, the number of direction dependent effective gains (i.e.\ source clusters) should be $Q < P(P-1)/(2P - 1) \approx (P-1)/2$, i.e. less than half of the number of stations $P$. 
	\end{itemize}
	Using $K$ frequency channels and forcing the gains to be polynomial of order $n$, we can similarly show that $Q < (K/(n+1))(P-1)/2$. So the number of clusters could be increased by a factor $K/(n+1)$. Hence
	\begin{itemize}
		\item \sl For multi frequency channel calibration, the number of direction dependent effective gains (i.e.\ source clusters) should be less than $Q < (K/(n+1))(P-1)/2$, for $P$ stations, $K$ channels, and $n$ parameters of the frequency dependent gain model. 
	\end{itemize}
	Taking LOFAR as an example, $K\approx 200$ for a 40 MHz bandwidth and taking a third-order polynomial we can theoretically increase the number of clusters by a factor 50 or approximately $Q < 25 P$ where $P$ is the number of station being used. In practice we should choose $Q$ to be considerably smaller to avoid over-fitting to noise and numerical problems. This bound becomes smaller if we introduce a baseline cut because the number of measurements decreases.
	Analyzing this problem using the Jacobian, we show that there exists a closed form basis for the null space of $\bJ$ (assuming more than one source per cluster). Let us denote this basis by $\bZ$, such that we have
	\begin{equation}
	\label{eq:null_DISage}
	\bZ = \left[ \begin{array}{ccc}
	\bZ_1 & \zeros & \zeros\\
	\zeros & \ddots & \zeros \\ 
	\zeros & \zeros & \bZ_Q\\\hline
	&\zeros&
	\end{array}\right],
	\end{equation}
	where 
	\[
	\bZ_q=\left [ \begin{array}{c}
	\bg_q \\
	-\bg_q^*
	\end{array} \right].
	\]
	In order to show that this is indeed a basis for the null space of $\bJ$ we need to show that $\bJ\bZ = \zeros$ and $\rank(\bZ)=Q$. We have $\bJ\bZ = \sum_q \bJ_q\bZ_q$ and using \eqref{eq:null_space_J_DI} we know $\bJ_q\bZ_q = \zeros$ for all $q=1,\dots,Q$. Because $\bZ$ is block-diagonal, as long as $\bg_q \neq \zeros $ for all $q$, $\rank(\bZ) = Q$. This Jacobian is then rank deficient by at least $Q$ which means that as discussed above at least $Q$ additional constraints are needed.
	
	\subsubsection{Direction dependent hierarchical beam-formed gains including all frequency channels}
	
	Similar to the previous models, we will still assume that the sky model is adequate. If this is the case, the only degeneracy that this model suffers from is the gain ambiguity. Using the result from Sect.\ \ref{ss:multichan} we obtain
	\[
	\begin{bmatrix}
	\bg_1 & \bg_2& \dots & \bg_K 
	\end{bmatrix} = \begin{bmatrix}
	\bc_1 & \bc_2& \dots & \bc_n 
	\end{bmatrix}(\bV_K^n)^T
	\]
	where $\bc_i$ for $i = 1 , \dots, n$ are the coefficients of the polynomials in $\btheta_{\text{smooth}}$. In this case the null space of the Jacobian $\bJ \bVt$ is give by
	\[
	\bz = \begin{bmatrix}
	\bc \\
	-\bc^*
	\end{bmatrix},\text{ where } \bc = \begin{bmatrix} \bc_1 \\
	\vdots\\
	\bc_n\\
	\end{bmatrix}.
	\]
	Even though, the smoothness constraint in Eqn.\,\eqref{eq:constrainedtheta} reduces the number of unknown significantly (by a factor $K/n$), because it constrains the variables and their conjugate independently, it does not remove the phase ambiguity and the constraint Jacobian $\bJ \bVt$ is also rank deficient by one. This means that the polynomial coefficients also suffer from the phase ambiguity. 
	
	%\subsection{Regularization}
	%\label{ssec:regularization}
	\noindent
	Whereas in the above discussion the gain solutions (as function of frequency) follow a particular functional form (e.g.\ polynomial), in e.g.\ calibration model discussed by in \citet{Yatawatta2015}, at each iteration, the functional form acts as a prior, penalizing deviations which are not required by the data. This Bayesian approach prefers smooth solutions but not so smooth that the data can not be modeled anymore. 
	\noindent
	In \citet{MouriSardarabadi2018b} we have investigated what the impact of both exact constraints and regularization is on the ability to calibrate an array in multiple directions using the effective model. In this paper, in addition to doing the same for hierarchical model, we also discuss the effect of a baseline cut on the CRB and present a different way to specifically S/N.
	
	\section{Cram\'er--Rao Bound in Calibration}
	\label{sec:CRB}
	
	In this section we investigate what the impact of various
	calibration schemes is on the Cram\'er--Rao Bound of the unknown parameters in $\btheta$. The CRB for radio astronomical data processing has been studied by several authors in past \cite{Wijnholds2008,Trott2016,MouriSardarabadi2016a}. While other studies focus on the noise on the calibrating parameters, such as the gains, we are more interested in the theoretical noise on the (residual) desired signal or $\bR_r$. It is important to investigate how the choice of the baseline cut, basis functions, sky model, etc. affect the minimum theoretical noise on these estimates. 
	\subsection{Unpolarized Direction Independent Model}\label{sec:CRBDIUP}
	
	Given an estimate of $\btheta$ denoted by $\bthetah$ the covariance matrix of the noise on this estimator, $\bC=\Cov(\btheta-\bthetah)$, is bound from below by the CRB. For Gaussian distributed data 
	\begin{equation}
	\bC=\bF^{-1}(\btheta)=\frac{1}{N}[\bJ(\btheta)^H(\bR^{-T}(\btheta)\otimes\bR^{-1}(\btheta))\bJ(\btheta)]^{-1},
	\end{equation} 
	where $\bF$ is the Fisher information matrix (FIM) and
	\begin{equation}
	\bJ=\frac{\partial \vect(\bR)}{\partial \btheta^T}=\begin{bmatrix}
	\bJ_{\bg} & \bJ_{\bg^*}  & \bJ_{\bsigma_{ps}} & \bJ_{\bsigma_{\bn}}& \bJ_{r} 
	\end{bmatrix}
	\end{equation}
	is the Jacobian matrix of $\bR$.
	If there are any model ambiguities then the FIM becomes singular and additional constraints (e.g.\ regularization) should be applied which changes the definition of CRB. In our model there are two ambiguities depending on {\tt Scheme\,1 or 2}, as discussed in Sect.\ \ref{sec:identif}: one phase ambiguity on the gains (e.g. $\bg$ and $e^{j\phi}\bg$ are both valid solutions) and another scale ambiguity between the gains and the sources (e.g. if $\bg$ is a solution so is $\sqrt{\alpha} \bg$ because we can scale the rest of the unknowns with a factor $1/\alpha$). Hence the FIM is rank deficient by two and some constraints are needed. In \cite{Wijnholds2009,Wijnholds2010a}, the technique described by \cite{Jagannatham2004} was used to illustrate the effect of different constraints. We follow a similar approach here. Let us define the constraints as a set of functions. We stack these functions in a $2 \times 1$ vector function $\bh(\btheta)$ such that
	\begin{equation}
	\bh(\btheta)=\zeros.
	\end{equation}
	Furthermore, let 
	\[
	\bH(\btheta)=\frac{\partial \bh(\btheta)}{\partial \btheta^T}
	\]
	be the Jacobian of the constraint functions and $\bUt$ be a unitary matrix for the null space of $\bH$ such that $\bH\bUt=\zeros$. Then the CRB for the constraint problem becomes
	\begin{equation}
	\label{eq:CRB_U0}
	\bC=\bUt(\bUt^H\bF\bUt)^{-1}\bUt^H.
	\end{equation}
	We put only constraints on the gains and hence the Jacobian $\bH$ with respect to the rest of the parameters is zero. As a result $\bUt$ has the following structure
	\begin{equation}
	\bUt=\left [\begin{array}{c|c}
	\bU & \zeros \\ \hline
	\zeros & \bI
	\end{array}\right]
	\end{equation}
	and hence
	\begin{equation}
	\label{eq:CRB_U}
	\bC=\left [\begin{array}{c|c}
	\bU & \zeros \\ \hline
	\zeros & \bI
	\end{array}\right]\left(\left [\begin{array}{c|c}
	\bU^H & \zeros \\ \hline
	\zeros & \bI
	\end{array}\right]\bF\left [\begin{array}{c|c}
	\bU & \zeros \\ \hline
	\zeros & \bI
	\end{array}\right]\right)^{-1}\left [\begin{array}{c|c}
	\bU^H & \zeros \\ \hline
	\zeros & \bI
	\end{array}\right].
	\end{equation}
	We can also partition the FIM as
	\begin{equation}
	\label{eq:Fisher_Part}
	\bF=\left [\begin{array}{c|c}
	\bF_{cc} & \bF_{cr} \\ \hline
	\bF_{cr}^H & \bF_{rr}
	\end{array}\right],
	\end{equation}
	where
	\begin{align}
	\bF_{cc}=N\bJ_{c}^H(\bR^{-T}\otimes\bR^{-1})\bJ_{c} \\
	\bF_{cr}=N\bJ_{c}^H(\bR^{-T}\otimes\bR^{-1})\bJ_{r} \\
	\bF_{rr}=N\bJ_{r}^H(\bR^{-T}\otimes\bR^{-1})\bJ_{r} 
	\end{align}
	and 
	\begin{equation}
	\bJ_c=\begin{bmatrix}
	\bJ_{\bg} & \bJ_{\bg^*}  & \bJ_{\bsigma_{ps}} & \bJ_{\bsigma_{\bn}}
	\end{bmatrix}
	\end{equation}
	is the Jacobian with respect to calibration parameters. We can combine \eqref{eq:CRB_U} and \eqref{eq:Fisher_Part} which leads to
	
	\begin{equation}
	\label{eq:CRB_U2}
	\bC=\left [\begin{array}{c|c}
	\bU & \zeros \\ \hline
	\zeros & \bI
	\end{array}\right]\left(\left [\begin{array}{c|c}
	\bU^H\bF_{cc}\bU & \bU^H\bF_{cr} \\ \hline
	\bF_{cr}^H\bU& \bF_{rr}
	\end{array}\right]\right)^{-1}\left [\begin{array}{c|c}
	\bU^H & \zeros \\ \hline
	\zeros & \bI
	\end{array}\right].
	\end{equation}
	We are interested in $\bC_{rr}$ which is the covariance matrix of the noise on the estimator of the extended emissions. Using matrix inverse lemma we have
	\begin{equation}
	\label{CRB_CRR}
	\begin{array}{l}
	\bC_{rr}=\bF_{rr}^{-1}\notag \vspace{1mm} \\
	+\bF_{rr}^{-1}\bF_{cr}^H\bU(\bU^H\bF_{cc}\bU-\bU^H\bF_{cr}\bF_{rr}^{-1}\bF_{cr}^H\bU)^{-1}\bU^H\bF_{cr} \bF_{rr}^{-1}.
	\end{array}
	\end{equation}
	Hence the noise on the shorter baselines is the sum of two uncorrelated noise contributions. The first one is the noise that we would have on the shorter baselines even if we would know the gains perfectly, and the second noise is the excess noise due to calibration (with a cut) denoted by $\bn_{exc}$ which has a complex Gaussian distribution $\MCN(\zeros,\bC_{exc})$ where 
	\begin{equation}
	\begin{array}{l}
	\bC_{exc}=\\\bF_{rr}^{-1}\bF_{cr}^H\bU(\bU^H\bF_{cc}\bU-\bU^H\bF_{cr}\bF_{rr}^{-1}\bF_{cr}^H\bU)^{-1}\bU^H\bF_{cr} \bF_{rr}^{-1}.
	\end{array}
	\end{equation}
	Analyzing this result is complicated due to its dependency on the particular constraint used during the implementation of the calibration algorithm. However, \citet{Jagannatham2004} shows that the Moore--Penrose pseudo-inverse of the unconstrained FIM corresponds to the best regularization in terms of the total variance of the estimates (i.e. sum of the variances for all of the estimated parameters is minimized). So we can define the covariance for the best possible excess noise as 
	\begin{equation}
	\label{eq:pinvFExcess}
	\bC_{exc}= [\bF^\dagger]_{rr}-\bF_{rr}^{-1},
	\end{equation}
	where $^\dagger$ is the Moore--Penrose pseudoinverse and $[\bF^\dagger]_{rr}$ is the submatrix of $\bF^\dagger$ corresponding to $\bsigma_{r}$. This measure of excess noise is independent of the choice of constraints and depends only on the model used. This allows us to analyze the problem without the need of taking any particular regularization into account.
	For an identifiable problem, i.e. one where $[\bJ_{\bg},\bJ_{\bg^*}]$ and $\bJ_{\bsigma_r}$ are linearly independent, we can show that (see Appendix\ \ref{app:prove_blockpinv})
	\begin{equation}
	[\bF^\dagger]_{rr} = \bF_{rr}^{-1}+\bF_{rr}^{-1}\bF_{cr}^H(\bF_{cc}-\bF_{cr}\bF_{rr}^{-1}\bF_{cr}^H)^\dagger\bF_{cr}\bF_{rr}^{-1},
	\end{equation}
	and
	\begin{equation}
	\label{eq:excess_noise_covariance}
	\bC_{exc} = \bF_{rr}^{-1}\bF_{cr}^H(\bF_{cc}-\bF_{cr}\bF_{rr}^{-1}\bF_{cr}^H)^\dagger\bF_{cr}\bF_{rr}^{-1}.
	\end{equation}
	In the multi-channel and beam-formed scheme the expression for the Jacobian and the FIM differs slightly but in Appendix\ \ref{app:MCMTCRB} we show that the excess noise can still be calculated using \eqref{eq:excess_noise_covariance}. The CRB for multi-channel and beam-formed gains are summarized by relations \eqref{eq:Frrmultibeam} and \eqref{eq:BLOCINVQ}, where $\bv_k$ is the $k$th row of $\bV_K^n$ (stacked into a column).
	\begin{table*}
		\begin{tabular}{c}
			%\hline \\
			\vbox{
				\begin{align}
				\label{eq:Frrmultibeam}
				\bF_{rr} = \bdiag(\bF_{\bsigma_r\bsigma_r,t,k})  = \begin{bmatrix}
				\bF_{\bsigma_r\bsigma_r,1,1} & & & & \\
				& \ddots && & \\
				&& \bF_{\bsigma_r\bsigma_r,T,1} &&\\
				&&& \ddots & \\
				&&&& \bF_{\bsigma_r\bsigma_r,T,K}
				\end{bmatrix},
				\end{align}
				\begin{equation}
				\label{eq:BLOCINVQ}
				\begin{array}{l}
				\bF_{cc}-\bF_{cr}\bF_{rr}^{-1}\bF_{cr}^H = \\
				\sum_{k=1}^K \begin{bmatrix}
				\bv_k\bv_k^T \otimes \sum_{t=1}^T (\bF_{\bg\bg}-\bF_{\bg\bsigma_r}\bF_{\bsigma_r\bsigma_r}^{-1}\bF_{\bg\bsigma_r}^H)_{k,t} & \bv_k\bv_k^T \otimes \sum_{t=1}^T (\bF_{\bg\bg^*}-\bF_{\bg\bsigma_r}\bF_{\bsigma_r\bsigma_r}^{-1}\bF_{\bg^*\bsigma_r}^H)_{k,t} \\
				\left ( \bv_k\bv_k^T \otimes \sum_{t=1}^T (\bF_{\bg\bg^*}-\bF_{\bg\bsigma_r}\bF_{\bsigma_r\bsigma_r}^{-1}\bF_{\bg^*\bsigma_r}^H)_{k,t} \right)^H & \bv_k\bv_k^T \otimes \sum_{t=1}^T (\bF_{\bg^*\bg^*}-\bF_{\bg^*\bsigma_r}\bF_{\bsigma_r\bsigma_r}^{-1}\bF_{\bg^*\bsigma_r}^H)_{k,t}
				\end{bmatrix}.
				\end{array}
				\end{equation}
			}
			%\hline
		\end{tabular}
	\end{table*} 
	These relations show that the CRB for the entire dataset can be calculated by combining the results for each frequency channel separately. This makes parallel computation possible. Because $\bF_{rr}$ is block diagonal, the only place where the smooth model, the duration of the stability of the gains and the number of channels come into play is in \eqref{eq:BLOCINVQ}, which also defines the excess noise. Showing that the excess noise strongly depends on the model assumptions. 
	
	\subsection{Fisher-Bayes Bound}
	One problem with the regularization scheme discussed in \cite{MouriSardarabadi2018b} is the difficulty to device theoretical bounds on the variance of the estimates. As the regularization method described is closely related to a Bayesian 
	with $\bC_{\btheta_g} = \Cov(\btheta_g)$ is the prior covariance of the gains. For this model we can use the Fisher-Bayes bound \citep[171-174]{Schreier2010}. The bound is then given by
	\[
	\bC = \left (\bF + \bC_{\btheta}^\dagger \right)^{-1}
	\]
	where the Fisher information matrix is calculated with respect to the original unknowns without any regularization or prior. Using the results presenter in  \citet{MouriSardarabadi2018b}, a simple prior for the polynomial smooth model can be written as $\bC_{\btheta} = \sigma^2 (\bP_{\bVt}^\bot \otimes \bI) $, where
	\[
	\bP^\bot_{\bVt} = \bI - \bVt\bVt^\dagger = \bU_n\bU_n^H,
	\]
	and $\bU_n$ is a unitary basis for the null space of $\bVt$. Using $\bU_n$ we can transform the unknown parameters in such a way that all pseudo inverses can be reduced to exact inverses. However, in order to keep some level of physical interpretation for the results, we do not change the parameter space here. The effect of this extra extra positive (semi-) definite (PSD) matrix to the Fisher, leads to a lower variance,as the inverse of the sum of two PSD matrices is smaller than the inverse of the individual matrices (i.e. $\bA^{-1}- (\bA+\bB)^{-1}$ is a PSD matrix). 
	
	\iffalse
	\subsection{Cram\'er--Rao Bound for Polarized Model}
	In this section we derive the CRB for Polarized model. We assume that the elements in $\bJt_{ps}$ are estimated directly from data and that the elements of $\bJt_r$ are estimated as a function of $\bJt_{ps}$ (i.e. we use $\bJt_r(\bJt_{ps})$).
	
	Following the same procedure as in previous section we can show that again we can decompose the variance on the estimates for $\bRt_r$ into two matrices, one that does not depend on direction independent gains and another term which does. However, this time both terms depend on the direction dependent effects (DDE) and hence the definition of excess noise is not readily given. 
	Another difficulty is that we need to calculate
	\[
	\frac{\partial \vect(\bJt_r)}{\partial [\vect^T(\bJt_{ps}) ,~\vect^H(\bJt_{ps})]}
	\]
	which causes the CRB to depend on the interpolation function used.
	\fi
	\section{Model Incompleteness}
	\label{sec:incomp}
	
	In any learning process the currently available data model is subject to errors and various types of incompleteness. These types of systematic errors lead to a biased estimation of the unknown parameters. In this section we give first and second order approximations for the bias as a result of incomplete deterministic calibration models. An example of an incomplete deterministic calibration model is excluding part of the available data model from calibration and studying its effect on the final results. In this case the incompleteness is known and fix.
	
	\noindent
	For simplification let us assume the following Least Squares (LS) problem
	\[
	\bthetah = \arg \min_{\btheta} \| \bR(\btheta_0) - \bR_0(\btheta) \|_F^2
	\]
	where $\bR(\btheta)= \bR_0(\btheta) + \bR_1(\btheta)$. It is then clear that the data model is incomplete because the model does not include $\bR_1(\btheta)$. Considering that the problem is non-linear we use a first order approximation of the solution. In such analysis we assume that the bias introduced is relatively small. It is also important to note that we use the exact noise-free $\bR(\btheta)$ and hence the results are asymptotic for $N \to \infty$. This is justifiable because we are interested in systematic bias errors which do not decrease with sample size.
	
	As stated above we assume that the bias is relatively small and hence if we start the non-linear search for the solution using a gradient descend we end up very closely to the solution of our LS problem. This leads to the following relation between the true solution and the biased one
	\[
	\begin{array}{rl}
	\bthetah &= \btheta_0 + \mu \bJ(\btheta_0)^H\vect\left[\bR(\btheta_0)-\bR_0(\btheta_0)\right ]\\
	&=\btheta_0 + \mu \bJ(\btheta_0)^H\vect\left [\bR_1(\btheta_0)\right]
	\end{array}
	\]
	where $\mu$ is a small positive constant and the bias is \mbox{$\bb(\btheta)= \mu \bJ(\btheta)^H\vect\left [\bR_1(\btheta)\right]$}. Let $\bC$ be the Cram\'er--Rao bound as derived above for the unbiased estimator of $\btheta$, using this bias term the biased CRB is now given by \citep[pp 27-77]{Kay1993a}
	\begin{equation}
	\label{eq:biasedcrb}
	\bC_{biased} = \left [ \bI + \frac{\partial \bb(\btheta)}{\partial \btheta^T}\right]\bC\left(\bR(\btheta_0)\right) \left [\bI + \frac{\partial \bb(\btheta)}{\partial \btheta^T} \right ]^H.
	\end{equation}
	If the bias is too large for a first order approximation a second order approximation can be used. In this case we have the following system of equations for the bias
	\[
	\bH_{\text{LS}}(\btheta) \bb(\btheta)= \bJ(\btheta)^H\vect\left [\bR_1(\btheta)\right]
	\]
	where $\bH_{\text{LS}}(\btheta)$ is the Hessian of the LS cost function which can be approximated as $\bJ(\btheta)^H\bJ(\btheta)$ for most practical purposes (this approximation is the core idea behind Gauss--Newton optimization processes).

	\section{Semi-Linearity, Excess Noise and Suppression}
	\label{sec:SL_EN_SP}
	
	In this section we show that the models presented above posses a property we call semi-linearity, introduced in \cite{MouriSardarabadi2018b}, to discuss bias-variance trade off which results from a baseline cut. We call a vector function semi-linear if $\bff(\btheta) = \bJ(\btheta)\bM\btheta$ where $\bM$ is a fixed matrix and $\bJ(\btheta) = \partial \bff(\btheta) / \partial \btheta^T$. This is closely related to solving a system of multivariate polynomials and is extensively studied within the field of Algebraic Geometry (AG). However, the study of calibratability using AG is beyond the scope of this paper and will be addressed in the future. 
	
	It is easy to verify that all the gain models introduced in the previous sections have this property. In this section we use the least squares (LS) cost function as an example to analyze the residuals and the bias-variance trade of calibrating with and without a cut. The gradient of LS cost function, denoted by $\bgamma$, is zero at the solution and hence we have
	\[
	\bgamma = \bJ^H\left (\brh-\bJ(\bthetah)\bM\bthetah \right)=\zeros,
	\]
	where $\bthetah$ is the LS estimator for $\btheta$.
	This leads to the following relation at the solution
	\[
	\bM \bthetah = \bJ^\dagger \brh.
	\]
	Using this relation we have for the residuals, denoted by $\bd$,
	\[
	\bd = \brh - \br(\btheta) = \brh - \bJ(\bthetah)\bJ(\bthetah)^\dagger\brh =(\bI - \bP(\bthetah))\brh ,
	\]
	where $\bP(\bthetah) =  \bJ(\bthetah)\bJ(\bthetah)^\dagger$ is a projection into the column space of $\bJ(\bthetah)$. For simplicity of the notation we define  \mbox{$\bPh^\bot:=\bI - \bP(\bthetah)$}. In practice $\brh$ contains contributions from sources that are not modeled and the desired signals such as EoR. Putting all these contributions into the model for $\brh$ the residuals become
	\[
	\bd = \bPh^\bot(\br + \br_{\text{EoR}} + \br_{\text{inc}} + \br_{\text{fg}} + \bepsilon )
	\] 
	where $\br$ is the part we have a model for and is used in the calibration, $\br_{\text{inc}}$ is the contribution of the unmodeled point sources, $\br_{\text{EoR}}$ is the EoR signal,  $\br_{\text{fg}}$ is the signal of unmodeled extended foregrounds and $\bepsilon$ is the thermal and finite sample noise. If we assume that the EoR signal is weak enough such that its contribution to the calibration results can be neglected, which is a reasonable assumption, then the projection matrix is completely defined by the model and the solution of the calibration algorithm. When the solutions are available from previous data analysis, the suppression of the a modeled EoR signal can be tested exactly without the need to rerun the calibration. Of course constructing the projection matrix in large setups could also be very expensive. For more comprehensive study of the residuals using semi-linearity we refer an interested reader to \cite{MouriSardarabadi2018b}.
	
	\noindent
	We can use the same principles to analyze the variance-bias trade off for a short baseline cut. In the notation that follows we use the subscript ${}_c$ to denote the calibrating parameters which are the gains if we do not have a cut. In this case we have
	\[
	\bthetah_c = 2\bJ_c^\dagger(\bthetah_c)\brh.
	\]
	The estimate for the shorter baselines becomes 
	\begin{align*}
	\bsigmah_{r,1} & = \bS^H (\brh - \frac{1}{2}\bJ_c(\bthetah_c)\bthetah_c)\\
	& = \bS^H\left(\bI-\bJ_c(\bthetah_c)\bJ_c^\dagger(\bthetah_c)\right)\brh\\
	& =\bS^H\bP_c^\bot(\bthetah_c)\brh
	\end{align*}
	where subscript $1$ is used for the case with no cut and $\bP_c^\bot$ is the projection into the null space of $\bJ_c$.
	
	Because the S/N of the model used for calibration is much higher than extended emission and the EoR we assume that $\bthetah_c$ and hence $\bP_c^\bot(\bthetah_c)$ are approximately the same in calibration with and without a cut. Using this approximation, the estimates for the shorter baselines with the cut can be shown  be
	\[
	\bsigmah_{r,2} \approx \left(\bS^H\bP_c^\bot(\bthetah_c)\bS \right)^{-1}\bS^H\bP_c^\bot(\bthetah_c)\brh.
	\]
	This shows that $\bsigmah_{r,2} \approx  \left(\bS^H\bP_c^\bot(\bthetah_c)\bS \right)^{-1} \bsigmah_{r,1}$.
	Now we can study the difference between these two solutions. Let us fill in the expression for $\brh$ in both cases. Doing so we have
	\[
	\bsigmah_{r,1} = \bS^H\bP_c^\bot(\bthetah_c) \left( \frac{1}{2}\bJ_c(\btheta_c) +  \bS\bsigma_r + \bepsilon \right)
	\]
	where we have ignored $\br_{\bn}$ because $\bS^H\br_{\bn} = \zeros$ and $\bepsilon$ is the noise on the visibilities. With sufficient S/N $\bthetah_c$ is a good estimate of $\btheta_c$ and as a result $\bP_c^\bot(\bthetah_c)\bJ_c(\btheta_c) \approx \zeros$ which leads to
	\begin{align*}
	\bsigmah_{r,1} &= \bS^H\bP_c^\bot(\bthetah_c)\bS\bsigma_r + \bS^H\bP_c^\bot(\bthetah_c)\bepsilon.
	\end{align*}
	We show that this solution is biased by calculating
	\begin{align*}
	\bb = \MCE\{\bsigmah_{r,1} - \bsigma_r \} &\approx  \bS^H\bP_c^\bot(\btheta_c)\bS\bsigma_r - \bsigma_r\\
	& = - \bS^H\left(\bI-\bP_c^\bot(\btheta_c)\right)\bS\bsigma_r.
	\end{align*}
	Because $\bP_c^\bot(\bthetah_c)$ is a projection matrix so is $\bI-\bP_c^\bot(\bthetah_c)$ and hence part (or all) of the extended emission and possibly EoR is subtracted and lost in this estimate. Experimental data supports this analysis \cite{Patil2016}. However, the noise on the data, $\bepsilon$, is multiplied by a projection matrix and a selection matrix and hence is reduced.
	
	Now we do the same for the solution with the cut where
	\[
	\begin{array}{rl}
	\bsigmah_{r,2} &\approx \left(\bS^H\bP_c^\bot(\bthetah_c)\bS \right)^{-1} \bsigmah_{r,1} \\
	&=  \bsigma_r + \left(\bS^H\bP_c^\bot(\bthetah_c)\bS \right)^{-1}\bS^H\bP_c^\bot(\bthetah_c)\bepsilon.
	\end{array}
	\]
	It is then clear that the estimate with the cut is unbiased however the noise term is multiplied by $\bHt^{-1}$ where $\bHt = \bS^H\bP_c^\bot(\bthetah_c)\bS$. Let the minimum eigenvalue of $\bHt$ be $\beta$. Because this matrix is a product of a selection matrix and a projection matrix $\beta \leq 1 $ and in worst case scheme the variance of the noise is increased by a factor $\propto 1/\beta$. The increase of the noise with respect to calibration without the cut is called the excess noise. 
	
	\noindent
	The minimum eigenvalue of $\bHt$ is also a measure of calibratability. If $\beta$ is zero then $\bJ_c$ and $\bJ_r = \bS$ are linearly dependent and hence by our definition of identifiability the problem is not calibratable. Hence, the excess noise increases as the problem becomes less identifiable. This clearly shows that it is not only the rank but also the condition number of $\bHt$ that is important for calibratability.  The condition number is a measure of precision loss, and a large condition number leads to sensitivity to noise (or amplification). Thus it possible for a problem to be theoretically identifiable but the condition number is too large with respect to the noise on the data and the parameters cannot be estimated. This is the problem of estimability and is related to the available S/N. In the next section we introduce a measure for signal to noise which in combination with the condition number should give us a better measure of calibratability.

	\subsection*{S/N per parameter}
	In this section we define a measure of available S/N per parameter. We use a definition of S/N which is related to the coefficient of variation \citep{Taguchi1986} and suggest the following multivariate version of this quanitiy:
	\[
	\text{S/N} = \diag(\bSigma_{\bx})^{-1/2} \bmu
	\]
	where $\bx$ is a vector of random variables with expected value $\bmu = \MCE\{\bx\}$ and covariance matrix $\bSigma_{\bx}$. Of course this is only valid for a non-zero mean variable.  To put this in perspective, by this definition an S/N of 19.6 for a Gaussian distributed random variable would mean that with a probability of 95\% the relative error is less than 10\%.
	
	\noindent
	To illustrate we look at the available S/N for a single visibility for a simple case where the array is only exposed to a single source at the center of the observing field and affected by white noise. Our random variable is $\brh = \vect(\bRh)$ with expected value $\bR = \sigma^2 \ones \ones^H + \sigma_n^2 \bI$. From \cite{Schreier2010} we have $\Cov(\brh) = (\bR^T \otimes \bR)/N$ and hence
	\[
	\begin{array}{rl}
	\text{S/N}_{\text{vis}} &= \sqrt{N} \diag(\bR^T \otimes \bR)^{-1/2} \br \\
	&= \sqrt{N} \vect\left[\diag(\bR)^{-1/2} \bR \diag(\bR)^{-1/2}\right]
	\end{array}
	\]
	which is a normalized covariance matrix (i.e. a correlation matrix). The off-diagonal elements are $\sqrt{N} \sigma^2 / (\sigma^2 + \sigma_n^2)$ which is a well known quantity for the available S/N on visibilities \citep{Briggs1995}.
	
	\noindent
	Using this definition and the semi-linearity introduced in the previous section, we can define the available S/N for a model as
	%\[
	%\text{S/N}_{\btheta} = \sqrt{N} \diag[(\bM^H\bJ^H(\bR^{-T} \otimes \bR^{-1})\bJ\bM)^\dagger]^{-1/2} \btheta.
	%\]
	\[
	\text{S/N}_{\btheta} = \sqrt{N} \diag[(\bJ^H(\bR^{-T} \otimes \bR^{-1})\bJ)^\dagger]^{-1/2} \btheta.
	\]
	For a given model we can calculate the ``instantaneous" S/N and deduce based on the average S/N or the worst case scheme how stable the instrument must be to have adequate gain solutions. This leads to extra constraints on the calibratability of an instrument. 
	
	\section{Simulations}
	\label{sec:SIM}
	
	\noindent
	In this section we  use simulated data and apply the theory presented in the paper to analyze calibratability, effect of polynomial smoothing on the gain solutions, calibration noise, excess noise as the result of the base line cut, gain model errors and sky model incompleteness. We are not simulating ionospheric effects and sky model errors. We also show how the introduce coefficient of variation can be used to find S/N per gain for different stations and clusters.
	
	\noindent
	For these simulation we use the LOFAR HBA configuration \citep{Haarlem2013}. We simulate 12 hours of data with solution intervals of 10s. The covariance matrices are simulated for $K=52$ channels with a 195.3 kHz each and integrated over 1 second which makes the total number of snapshots per solutions $T=600$. For LOFAR HBA we have a two level beam-forming hierarchy. Sky model is simulated on the lowest level and then software beam-formers and correlators are used to generate the reduced dataset at higher levels. Using the beamfomring model in Sect. \ref{ssec:cov_beam_data} we first beam-form 16x16 HBA tiles, then apply a single gain per tile and proceed with beam-forming the titles to generate the station output. The combination of tile beam, gains and station level correlations result in direction dependent beam fluctuations. As discussed in Sect.~\ref{sec:CRB} the covariance matrix of the noise on the data and the asymptotic noise on the estimates is completely defined by the noise free covariance matrices and the Jacobians. Using this we can generate noise with correct statistical properties for both the visibility (covariance) data as well as for the solutions at any stage post data generation.
	
	\noindent
	The sky model used consists of 20000+ components modeling the NCP field which is one of the fields currently used for EoR related studies \cite{Patil2017}. We additionally simulate the extended foreground emission as a random Gaussian field. The visibilities for this field follow a power law as function of baseline. This source is not included in the sky model and its visibilities on the shorter baselines are estimated using a cut.
	
	\noindent
	First we will discuss the effect of having a baseline cut for a complete sky-model. We then proceed and add sky-incompleteness to the simulations and study the bias.
	
	\subsection{Baseline cut, excess noise and sky incompleteness}
	We have already derived the expression for the excess noise in Sect.\ \ref{sec:CRB} and we know it must show frequency dependent behavior. We use the simulation setup which is described in previous section to generate noise with the same covariance structure as the CRB for both the scheme with and without the cut.

\begin{figure*}
		\subfloat[CRB bound for noise on shorter baselines with no cut\label{fig:noise_spectrum}]{%
			\includegraphics[width = 0.45\textwidth]{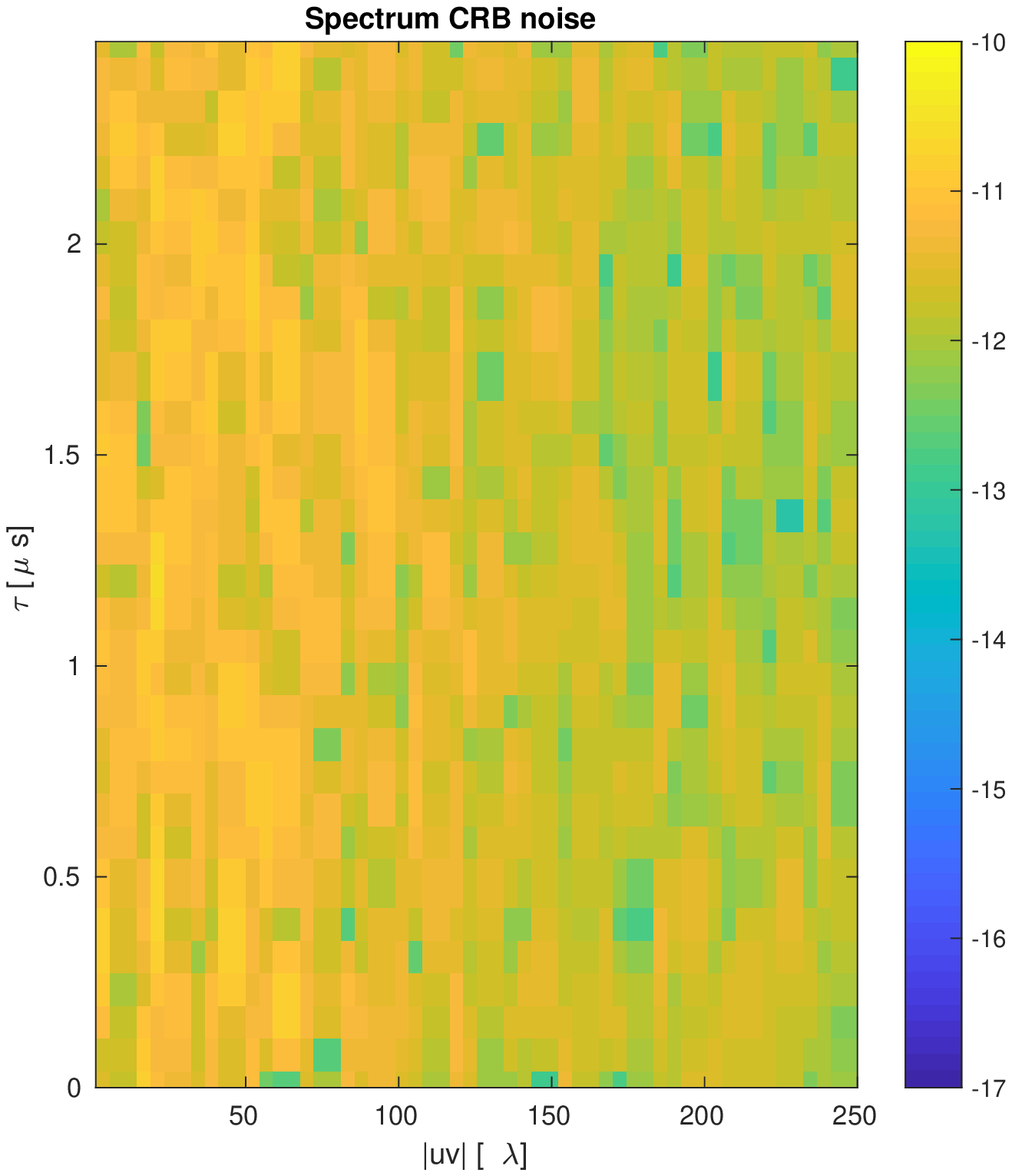}
		}
		\hfill
		\subfloat[CRB bound for excess noise on shorter baselines with cut\label{fig:excess_noise_spectrum}]{%
			\includegraphics[width = 0.45\textwidth]{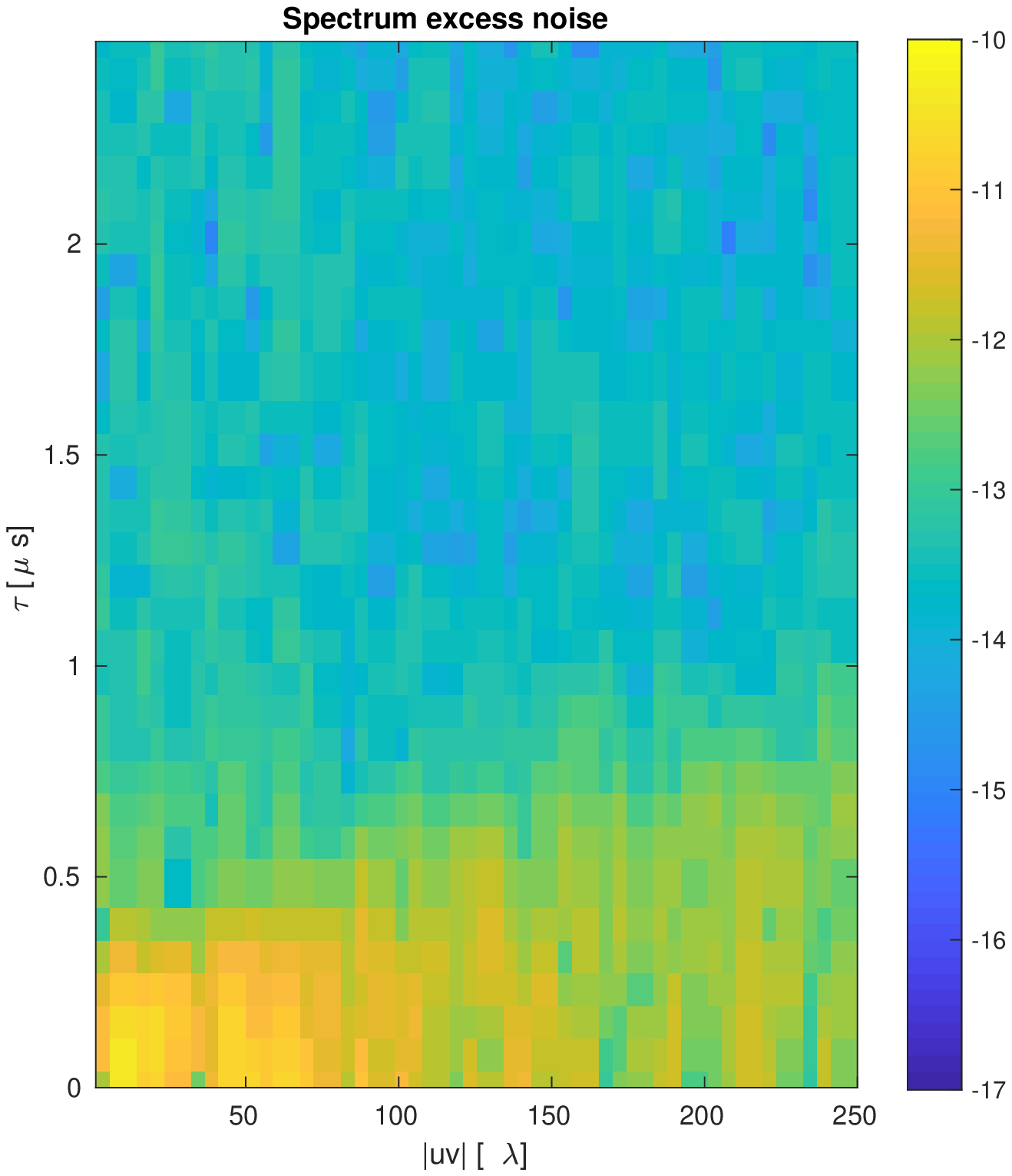}
		}
		\hfill
		\subfloat[First Order LS bias\label{fig:bias_spectrum}]{%
			\includegraphics[width = 0.45\textwidth]{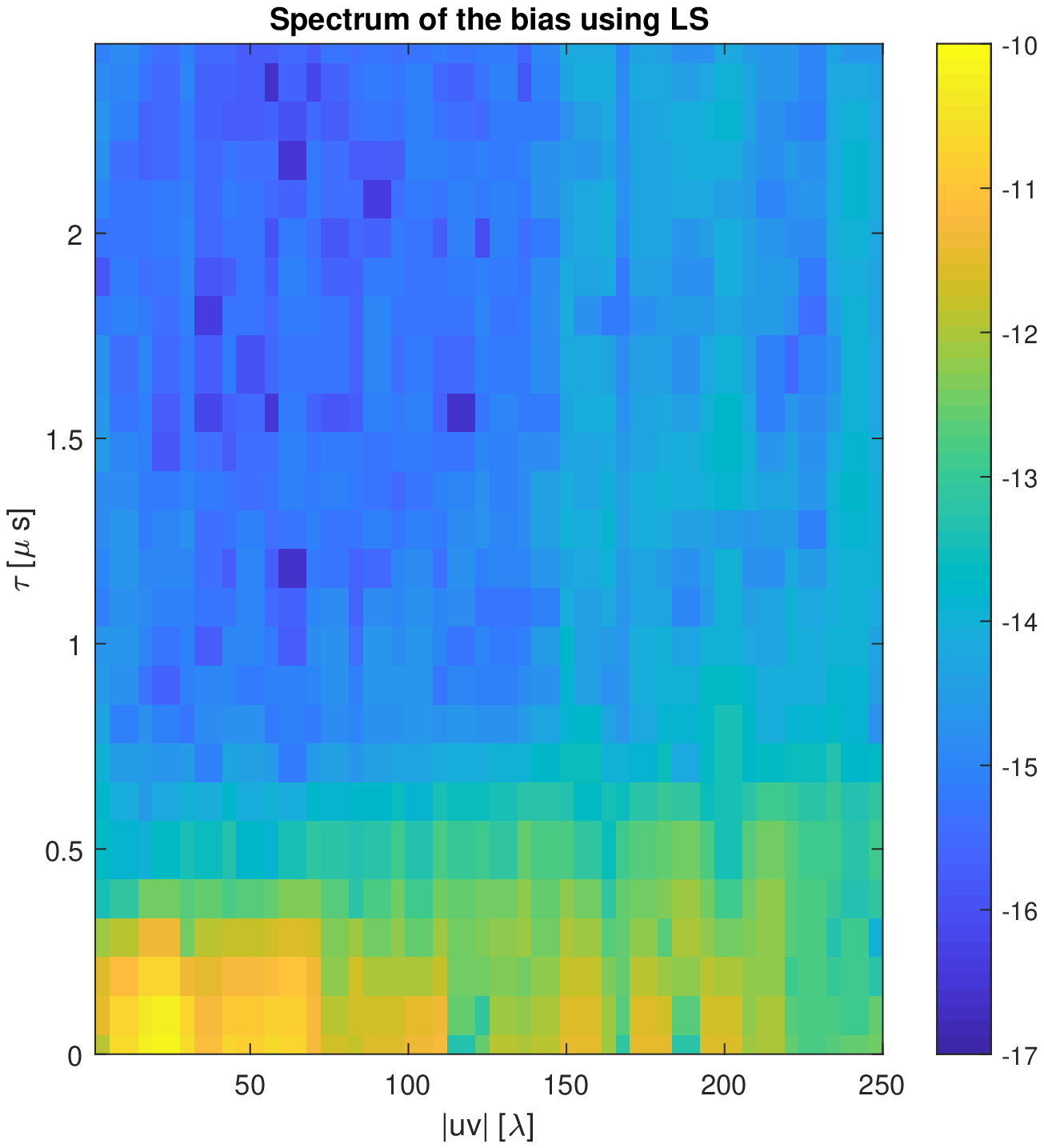}
		}
		\caption{}
		\label{fig:noise_and_excess_noise_spectrum}
	\end{figure*}

	\noindent
	Figure \ref{fig:noise_and_excess_noise_spectrum} illustrates the delay-baseline spectrum for each of the results. Figure\ \ref{fig:noise_spectrum} shows the spectrum for the CRB if we would know the gains exactly and estimate the shorter baselines using a maximum likelihood estimator. Figure\ \ref{fig:excess_noise_spectrum} shows the increase in the noise as the result of calibration. This is the theoretical effect of estimating the gain for each tile within each station. We clearly see that the spectrum of the excess noise shows frequency dependency and a wedge like feature. 
	
	\noindent
	Figure\ \ref{fig:bias_spectrum} shows the spectrum of the least squares (LS) bias when sky incompleteness is introduced. In order to simulate the sky incompleteness, all the model components with a flux less than 1 mJy where removed from estimation model. Surprisingly the effect of the bias above the wedge seem to be very limited. However, as the simulation currently does not include any ionospheric effects, this bias does not seem to average out much inside the beam.
	
	\begin{figure}[H]
		\includegraphics[width = 0.45\textwidth, height = 0.25\textheight]{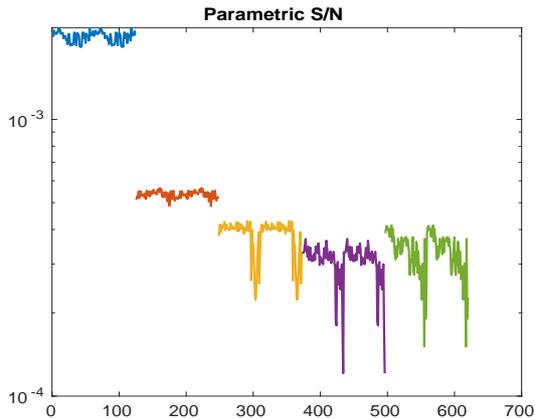}
		\caption{The instantaneous S/N per parameter for 5 clusters within the LOFAR NCP field. Each color indicates a different cluster. The figure shows the S/N for the gain parameters in $[\bg_q^T , \bg_q^H]$ for each cluster.}
		\label{fig:SNR}
	\end{figure}
	\subsection{Parametric S/N}
	In order to illustrate how the parametric S/N can reveal the quality of the sky model, we use the effective gains model to calculate the S/N per station and per direction. In order to do so we use the sky model used to calibrate the LOFAR telescope as described by \citet{Patil2017}. From this sky we use the first 5 brightest clusters (total flux in the cluster) and calculate their instantaneous S/N. Figure~\ref{fig:SNR} shows the results.
	
	\noindent
	Based on these result we see that while the S/N of the gains for all station are almost equal for the first two clusters, the S/N drops rapidly as we move from LOFAR core stations to LOFAR remote stations \cite{Haarlem2013} which have larger baselines, showing a baseline dependency for S/N. This is consistent with instantaneous $uvw$ converge of the HBA stations. Note that the sampling frequency for this observation mode is $200$ MHz and the S/N increases with $\sqrt{N}$. It is then clear that even for bright clusters, sufficient integration time is needed to achieve acceptable S/N.
	
\begin{figure*}
	\subfloat[Noise increase 50km\label{fig:50cut}]{%
		\includegraphics[width=0.49\textwidth]{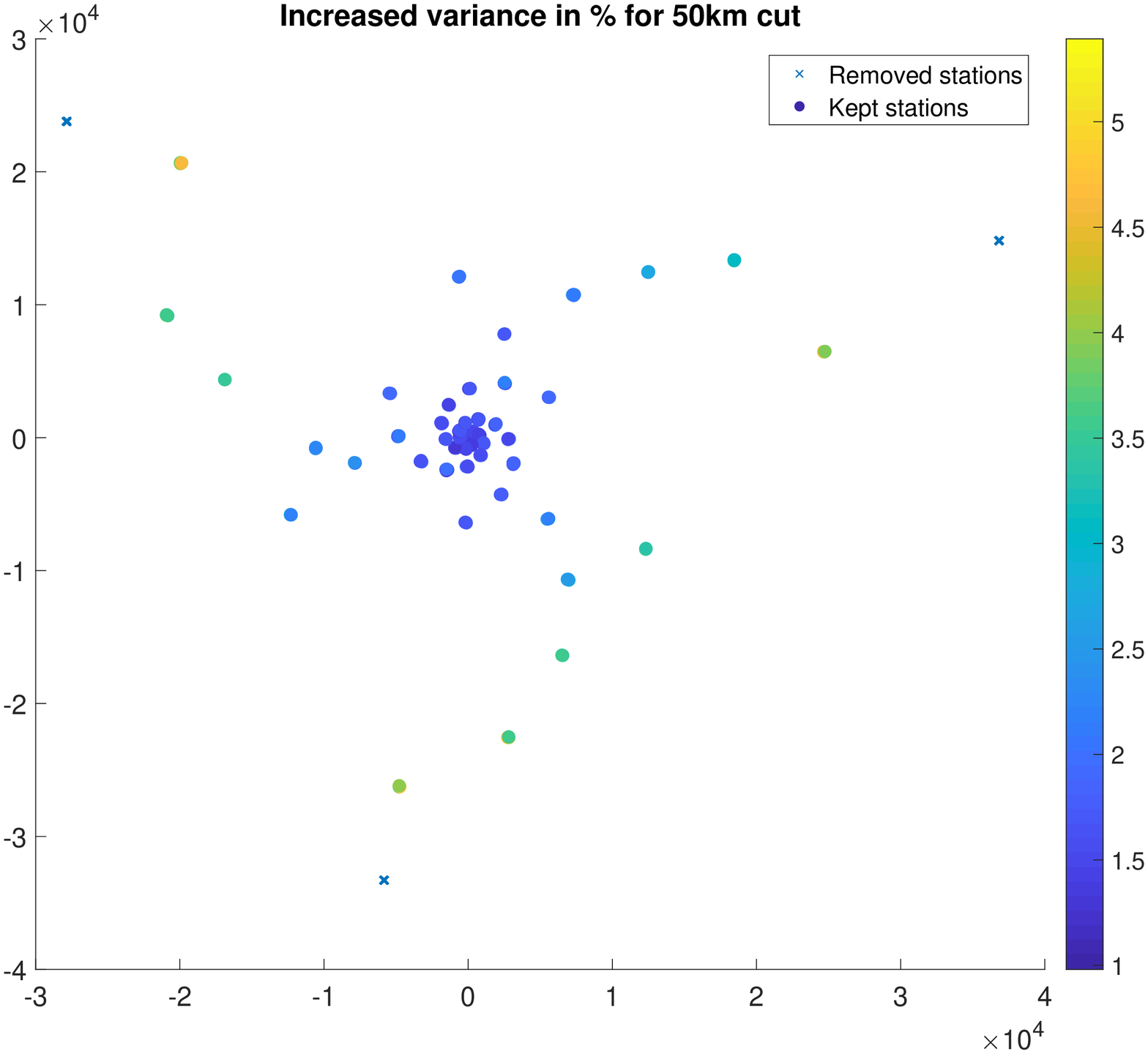}
	}
	\hfill
	\subfloat[Noise increase 40km\label{fig:40cut}]{%
		\includegraphics[width=0.49\textwidth]{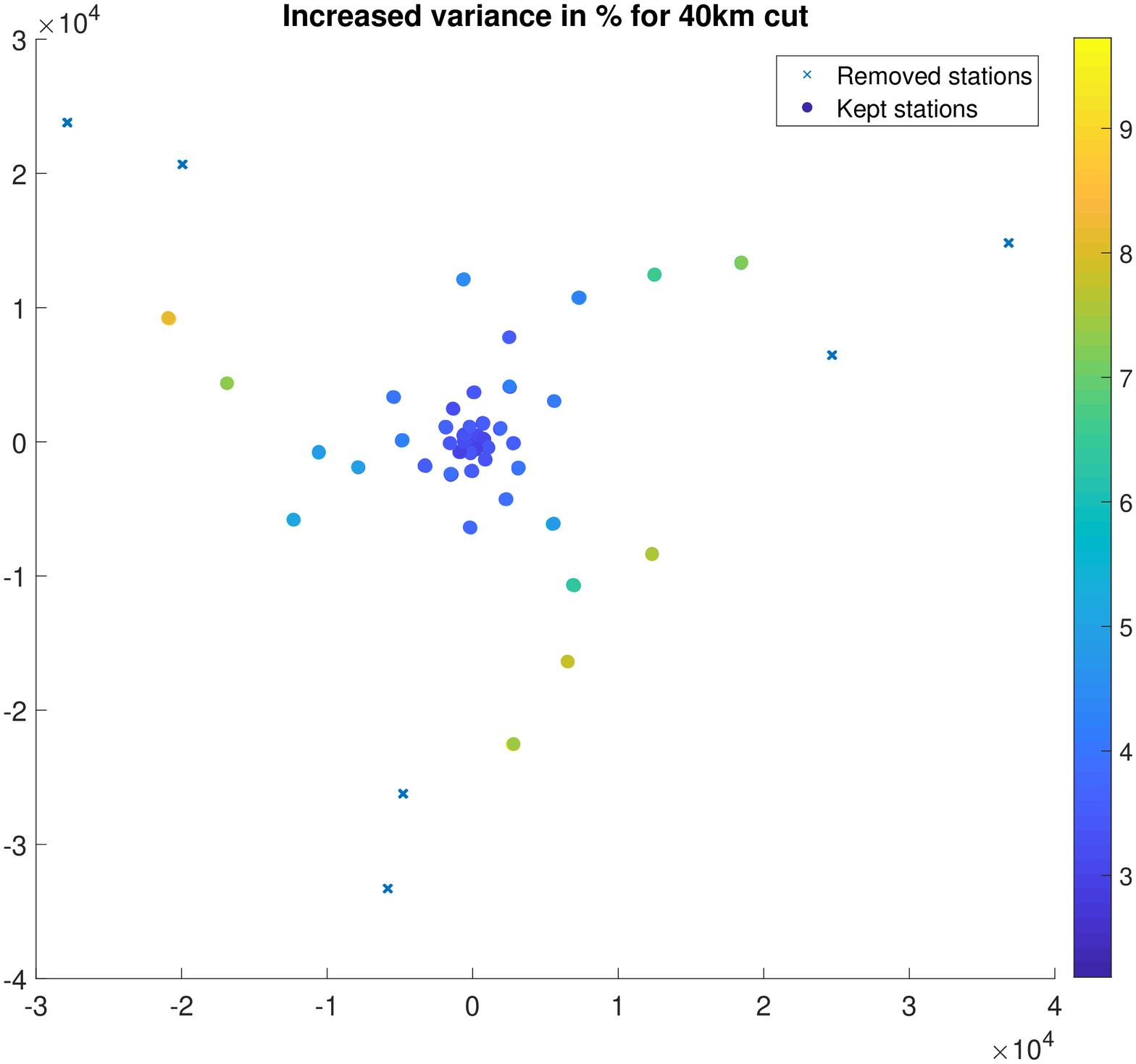}
	}
	
	\caption{}
	\label{fig:noise_and_excess_noise_spectrum}
\end{figure*}

\subsection{SKA Baseline cut}
In this section we show how an analysis based on the CRB can be used to study the effect of various design decisions. For this simulations we calculate the CRB for the gains solutions three times using the same sky model. We do this without a cut and we have a model for the diffused foreground following a power-law included in our sky model. The same NCP model used in the previous simulation is used for the point sources. We group the dipoles within each station based on their distance from each other and assume that each group shares the same gain. This effectively produces a tiled version of the SKA-LOW stations. We also assume that each tile has a smooth gain which can be modeled using a third order polynomial over a bandwidth of 10 Mhz.

Using the setup above, we calculate the variance of the noise on each tile based on the CRB. Figure\  \ref{fig:50cut} shows how much the CRB increases as we remove the outer stations in order to decrease the maximum baselines for the SKA-low to 50km.  Figure\ \ref{fig:40cut} shows the same results for a 40km maximum baseline. As is illustrated in the figures an increase of  $\approx 5\%$ and $\approx9\%$ on the noise variance can be expected for the 50km and 40km cut respectively.
	
	\section{Conclusions}
	In this paper we illustrated that the covariance model used for various gain calibration schemes is highly structured and by studying the Jacobian and the Fisher information matrix, we can gain valuable insights into the possible calibratability of a given instrument based on limited assumptions on the sky model. We illustrated how different constraints can be introduced and systematically studied with this approach.
	
	\noindent
	By introducing a model for each tile within a station which show smooth behaviors in frequency, we illustrated how different type of hard (forcing the smoothness by polynomials) and soft (using regularization) constraints can be analyzed and theoretical bounds can be calculated. These bounds we derived are a function of the measurement equation used for each calibration scheme and can be tested without a particular algorithm in mind. This also allows for design decisions which are not limited by algorithmic considerations.
	
	\noindent
	By showing the so called ``semi-linear" behavior of the calibration problem, we introduced a robust method to study the suppression problem and how an approach like a baseline cut affects it. The similarity of this method to the linear least squares method also allows for the insights and intuition learned from linear problems to be carried over to study the residuals in a calibration problem.
	
	\noindent
	In the future we will add the ionospheric effects to the covariance model in order to have a full theoretical description for any instrument from a signal processing perspective and to be able to find the theoretical bounds on the calibratability of future instruments.

	\bibliographystyle{mnras}
	\bibliography{biblio3}

	\appendix
	
	\section{Notations}\label{app:notation}
	
	A boldface letter such as $\bx$ denotes a column vector, a boldface
	capital letter such as $\bX$ denotes a matrix.
	$x_i$ is the $i$th element of the vector $\bx$.
	$\bI$ is an identity matrix of appropriate size and $\bI_p$ is a
	$p\times p$ identity matrix.
	
	$(\cdot)^T$ is the transpose operator, $(\cdot)^*$ is the complex
	conjugate operator, $(\cdot)^H$ is the Hermitian transpose,
	$\|\cdot\|_F$ is the Frobenius norm of a matrix, $\|.\|$ is the two
	norm of a vector, $\EExp\{\cdot\}$ is the expectation operator and $\MCN(\bmu,\bSigma_{\bs})$ represents the multivariate complex normal distribution with expected value $\bmu$ and covariance matrix $\bSigma_{\bs}$.
	
	$\vect(\cdot)$ stacks the columns of the argument matrix to form a
	vector, $\vectdiag(\cdot)$ stacks the diagonal elements of the argument
	matrix to form a vector, $\diag(\cdot)$ is a diagonal matrix with its
	diagonal entries from the argument vector (if the argument is a matrix
	$\diag(\cdot)=\diag(\vectdiag(\cdot))$).
	
	Let $\otimes$ denote the Kronecker product, i.e.,
	\[
	\bA \otimes \bB := 
	\left[\begin{array}{ccc}  
	a_{11} \bB & a_{12} \bB & \cdots\\
	a_{21} \bB & a_{22} \bB & \cdots\\
	\vdots     & \vdots     & \ddots \end{array} \right]
	\]
	Further, $\circ$ denotes the Khatri-Rao
	product (column-wise Kronecker product), i.e.,
	\[
	\bA \circ \bB := [\ba_1\otimes \bb_1, \ba_2 \otimes \bb_2 , \cdots]
	\]
	and $\odot$ denotes the Schur-Hadamard (element-wise) product.
	The following properties are used throughout the paper (for matrices and
	vectors with compatible dimensions):
	\begin{align}
	\label{kron:1} (\bB^T \otimes \bA)\vect(\bX)&=\vect(\bA\bX\bB) \\
	\label{kron:2} (\bB \otimes \bA)^H&=(\bB^H \otimes \bA^H)\\
	\label{kron:3} (\bB \otimes \bA)^{-1}&=(\bB^{-1} \otimes \bA^{-1})\\
	\label{kron:4} (\bB^T \circ \bA)\bx&=\vect(\bA\diag(\bx)\bB)\\
	\label{kron:5} (\bB\bC \otimes \bA\bD)&=(\bB \otimes \bA)(\bC \otimes \bD)\\
	\label{kron:6} (\bB\bC \circ \bA\bD)&=(\bB \otimes \bA)(\bC \circ \bD)\\
	\label{kron:7} (\bB^H\bC \odot \bA^H\bD)&=(\bB \circ \bA)^H(\bC \circ \bD)\\
	\label{kron:8} \vectdiag(\bA^H\bX\bA)&=(\bA^* \circ \bA)^H\vect(\bX)
	\end{align}
	Additionally for any $P \times Q$ matrix $\bA$ there exists a $PQ \times PQ$ permutation matrix $\bK_{P,Q}$ such that
	\begin{equation}
	\vect(\bA^T)=\bK_{P,Q}\vect(\bA).
	\end{equation}
	We also have $\bK^T\bK=\bI$, $\bK_{P,Q}=\bK_{Q,P}^T$ and hence $\bK_{P,Q}\bK_{Q,P}=\bI$.
	Using these relations, for any $P \times Q$ matrix $\bA$ and  $M \times N$ matrix $\bB$  we have
	\begin{align}
	\label{kron:11} (\bA \otimes \bB)\bK_{Q,N}&=\bK_{P,M}(\bB \otimes \bA) \\
	\label{kron:12} (\bA \circ \bB)&=\bK_{P,M}(\bB \circ\bA),
	\end{align} 
	where $Q = N$ for \eqref{kron:12}.
	
	\section{Block matrix Moore--Penrose pseudo--inverse}
	\label{app:prove_blockpinv}
	
	The pseudo inverse of a partitioned positive semi-definite matrix can be expressed in terms of the inverse and pseudo inverse of its submatrices \citep{Rohde1965}. In this section in order to simplify the notation we redefine 
	\[
	\bJ=(\bR^{-T/2} \otimes \bR^{-1/2})\frac{\partial \br}{\partial \btheta^T}
	\]
	which does not change the analysis because $\bR$ is non-singular.  For a positive semi-definite matrix of the form
	\[
	\bF = \begin{bmatrix}
	\bJ_c & \bJ_r
	\end{bmatrix}^H\begin{bmatrix}
	\bJ_c & \bJ_r
	\end{bmatrix} =  \begin{bmatrix}\bF_{cc} & \bF_{cr} \\ \bF_{cr}^H & \bF_{rr} \end{bmatrix} 
	\]
	where $\bF_{rr}$ is non-singular and \mbox{$\rank(\bF) = \rank(\bF_{cc}) + \rank(\bF{rr})$} we have \citep{Rohde1965}
	\[
	\bF^\dagger = \begin{bmatrix} \bF_{cc}^\dagger + \bF_{cc}^\dagger\bF_{cr}\bQ^{-1}\bF_{cr}^H\bF_{cc}^\dagger  & -\bF_{cc}^\dagger\bC\bQ^{-1} \\ -\bQ^{-1}\bF_{cr}^H\bF_{cc}^\dagger & \bQ^{-1} \end{bmatrix} 
	\]
	where \mbox{$\bQ = \bF_{rr} - \bF_{cr}^H \bF_{cc}^\dagger \bF_{cr}$} is invertible. The non-singularity of $\bF_{rr}$ is a necessary condition for identifiability of the calibration and must hold. This result provides closed from expression for the excess noise on the calibrating parameters, while we are interested in excess noise on the shorter base lines.
	
	For any Hermitian matrix $\bA$ we have \mbox{$(\bU^H\bA\bU)^\dagger = \bU^H\bA^\dagger\bU$ where $\bU$} is a unitary matrix. For the positive semi-definite matrix $\bF_{cc} =\bJ_c^H\bJ_c$ there always exists a unitary matrices such that $\bJ_c\bU = [\zeros, \bL]$ where $\bL$ has full column rank and
	\begin{equation}
	\bU^H\bF_{cc}\bU = \begin{bmatrix}
	\zeros & \zeros \\
	\zeros & \bL^H\bL
	\end{bmatrix}.
	\end{equation}
	This leads to
	\begin{equation}
	\bF_{cc}^\dagger = \bU \begin{bmatrix}
	\zeros & \zeros \\
	\zeros & (\bL^H\bL)^{-1}
	\end{bmatrix}\bU^H
	\end{equation}
	Following the same reasoning for $\bF$ we have
	\[
	\bUt^H\bF\bUt = \begin{bmatrix}
	\zeros & \zeros & \zeros\\
	\zeros & \bL^H\bL & \bL^H\bJ_r \\
	\zeros & \bJ_r^H \bL & \bF_{rr}
	\end{bmatrix}
	\]
	where 
	\[
	\bUt = \begin{bmatrix}
	\bU & \zeros \\
	\zeros & \bI
	\end{bmatrix}.
	\]
	Using this identity and the fact that both $\bF_{rr}$ and $\bL^H\bL$ are invertible, we have \mbox{$\bF^\dagger = \bUt(\bUt^H\bF\bUt)^\dagger\bUt^H$} and 
	\[
	\begin{array}{l}
	(\bUt^H\bF\bUt)^\dagger =\\  
	\begin{bmatrix}
	\zeros & \zeros & \zeros\\
	\zeros & \bQt^{-1} & -\bQt^{-1}\bL^H\bJ_r\bF_{rr}^{-1} \\
	\zeros & -\bF_{rr}^{-1}\bJ_r^H\bL \bQt^{-1} & \bF_{rr}^{-1} + \bF_{rr}^{-1}\bJ_r^H\bL\bQt^{-1}\bL^H\bJ_r\bF_{rr}^{-1}
	\end{bmatrix}\\
	\end{array}
	\]
	where $\bQt = \bL^H\bL - \bL^H\bJ_r\bF_{rr}^{-1}\bJ_r^H\bL$. Because of the structure of $\bUt$ the lower right block of $(\bUt^H\bF\bUt)^\dagger$ and $\bF^\dagger$ are the same and the closed form expression for the excess noise becomes
	\[
	\begin{array}{l}
	\bF_{rr}^{-1}\bJ_r^H\bL\bQt^{-1}\bL^H\bJ_r\bF_{rr}^{-1}  \\
	=\bF_{rr}^{-1}\bJ_r^H\begin{bmatrix}
	\zeros & \bL	
	\end{bmatrix}\begin{bmatrix}
	\zeros & \zeros \\
	\zeros & \bQt^{-1}
	\end{bmatrix}\begin{bmatrix}
	\zeros \\ \bL^H	
	\end{bmatrix}\bJ_r\bF_{rr}^{-1}\\
	= \bF_{rr}^{-1}\bJ_r^H\begin{bmatrix}
	\zeros & \bL	
	\end{bmatrix}\bU^H\bU\begin{bmatrix}
	\zeros & \zeros \\
	\zeros & \bQt^{-1}
	\end{bmatrix}\bU^H\bU\begin{bmatrix}
	\zeros \\ \bL^H	
	\end{bmatrix}\bJ_r\bF_{rr}^{-1}\\
	= \bF_{rr}^{-1}\bJ_r^H\bJ_c\left(\bU\begin{bmatrix}
	\zeros & \zeros \\
	\zeros & \bQt
	\end{bmatrix}\bU^H\right)^\dagger \bJ_c^H\bJ_r\bF_{rr}^{-1}\\
	=\bF_{rr}^{-1}\bF_{cr}^H(\bF_{cc}- \bF_{cr} \bF_{rr}^{-1} \bF_{cr}^H)^\dagger \bF_{cr}\bF_{rr}^{-1}.\\
	\end{array}
	\]
	This completes what needed to be shown.
	\section{Multi--channel and multi--snapshot CRB}
	\label{app:MCMTCRB}
	Because the signal on each frequency channel is statistically independent from another frequency the definition of CRB becomes
	\begin{align*}
	\bC=&\left (\sum_k^K \bF_k\right )^{-1}\\&
	=\frac{1}{N}\left [\sum_k^K \bJ_k^H(\bR_k^{-T}\otimes\bR_k^{-1})\bJ_k\right ]^{-1}\\
	&=\frac{1}{N}\left [ \bVt^H \bJ^H \bRt^{-1} \bJ\bVt \right ] ^{-1}
	\end{align*}
	where 
	\begin{equation}
	\bVt=\left[ \begin{array}{c|c}
	\bI_2 \otimes \bV_K^n \otimes \bI_{P}  & \zeros \\ \hline
	\zeros & \bI
	\end{array}\right ],
	\end{equation}
	\begin{equation}
	\bRt^{-1}=\left [\begin{array}{ccc}
	\bR_1^{-T}\otimes\bR_1^{-1}&  & \zeros  \\
	&\ddots & \\ 
	\zeros & & \bR_K^{-T}\otimes\bR_K^{-1}
	\end{array} \right],
	\end{equation}
	\begin{equation}
	\label{eq:JACOBIANS_MULTI_FREQ}
	\bJ=\left [\begin{array}{c|c}
	\bJ_{\bg}& \bJ_{\bsigma_{r}}
	\end{array} \right]
	\end{equation}
	\begin{equation}
	\label{eq:JACOBIANS_MULTI_FREQ_g}
	\bJ_{\bg}=\left [\begin{array}{c|c}
	\begin{array}{ccc}
	\bJ_{\bg_1} &  & \zeros  \\
	& \ddots &  \\ 
	\zeros &  & \bJ_{\bg_K}
	\end{array}
	&
	\begin{array}{ccc}
	\bJ_{\bg^*_1} &  & \zeros  \\
	& \ddots &  \\ 
	\zeros &  & \bJ_{\bg^*_K}
	\end{array}
	\end{array} \right]
	\end{equation}
	\begin{equation}
	\label{eq:JACOBIANS_MULTI_FREQ_sigma_r}
	\bJ_{\bsigma_{r}}=\left [\begin{array}{ccc}
	\bJ_{\bsigma_{r,1}} &  & \zeros  \\
	& \ddots &  \\ 
	\zeros &  & \bJ_{\bsigma_{r,K}}
	\end{array} \right]
	\end{equation}
	and we have
	\begin{align}
	\bJ_{\bg_k}&= \frac{\partial \vect(\bR_k)}{\partial \bg_k^T} = \bG_b^*\bR_{0,k}^T \circ \bB\\
	\bJ_{\bg^*_k}&= \frac{\partial \vect(\bR_k)}{\partial \bg_k^H} = \bB \circ \bG_b\bR_{0,k}\\
	\bJ_{\bsigma_{r,k}} & =\frac{\partial \vect(\bR_k)}{\partial \bsigma^T_{r,k}} = (\bG_b^*\bB^T\bP^{-1} \otimes \bG_b\bB^T\bP^{-1})\bS_k
	\end{align}
	where $\bR_{0,k}$ is defined using \eqref{eq:beam-formedR} as
	\[
	\bR_{0,k}=\bR_{ps}(\lambda_k)+ \bB^T\bP^{-1}\bR_r(\lambda_k)\bP^{-1}\bB
	\]
	and
	\[
	\vect(\bR_r(\lambda_k)) = \bS_k \bsigma_{r,K}.
	\]
	
	Using the block-diagonal structure of $\bJ$ and Kronecker structure of $\bVt$ we cam simplify the Jacobians in terms of the coefficients as
	\[
	\bJ_{\ba} = \begin{bmatrix}
	\bv_1^T \otimes \bJ_{\bg_{1,1}} \\
	\bv_1^T \otimes \bJ_{\bg_{1,2}} \\
	\vdots \\
	\bv_k^T \otimes \bJ_{\bg_{k,t}} \\
	\vdots \\
	\bv_K^T \otimes \bJ_{\bg_{K,T}} 
	\end{bmatrix}
	\]
	where $\bv_k$ is the $k$th row of $\bV_K^n$ stacked into a column (i.e. columns of ${\bV_K^n}^T$). The total Jacobian as a function of coefficients becomes
	\begin{equation}
	\label{eq:JACOBIANS_MULTI_FREQ2}
	\bJ(\ba,\bsigma_r)=\left [\begin{array}{c | c|c}
	\bJ_{\ba}& \bJ_{\ba^*}& \bJ_{\bsigma_{r}}
	\end{array} \right]
	\end{equation}
	\subsection*{Extension to Multiple time-snapshots}
	In many calibration schemes we assume that the gains are stable for a small period of time which can range from seconds to minutes. However given the resolution of the instrument the sky model could change in a faster rate which we call a time-snapshot. We assume that for snapshots $t=1,\dots, T$ the gains are stable. In this case we assume to have access to visibilities $\bR_{t,k}$ and hence we need only to extend the definition \eqref{eq:JACOBIANS_MULTI_FREQ} which leads to
	\begin{equation}
	\label{eq:JACOBIANS_MULTI_FREQ_MULTI_SNAPSHOT}
	\bJ_k=\left [\begin{array}{c|cccc }
	\bJ_{\bg_{k,1}}& \bJ_{\bsigma_{r,k,1}} & & & \zeros \\
	\bJ_{\bg_{k,2}}& & \bJ_{\bsigma_{r,k,2}} &  \\
	\vdots & & & \ddots & \\
	\bJ_{\bg_{k,1}}&\zeros& & &  \bJ_{\bsigma_{r,k,T}} 
	\end{array} \right]
	\end{equation}
	where ${}_t\bJ_{\bg}$ and ${}_t\bJ_{\bsigma_{r}}$ are defined in the same way as \eqref{eq:JACOBIANS_MULTI_FREQ_g} and \eqref{eq:JACOBIANS_MULTI_FREQ_sigma_r} respectively.
	
	Given the structure of $\bJ$ it is clear that the conditioning of this matrix increases as $T$ becomes larger, however there are many factors (such as temperature and humidity) which limit the stability of the instrument and hence $T$.
	
	Now we can give an expression for the FIM taking all aspects into account including the smoothness. 
	
	Let us define the submatrices for the fisher information without smoothness as
	\begin{align}
	\bF_{\bg\bg,t,k} & = \bJ_{\bg_{k,t}}^H(\bR_{t,k}^{-T} \otimes \bR_{t,k}^{-1})\bJ_{\bg_{k,t}}\\
	\bF_{\bg\bg^*,t,k}&= \bJ_{\bg_{k,t}}^H(\bR_{t,k}^{-T} \otimes \bR_{t,k}^{-1})\bJ_{\bg^*_{k,t}}\\
	\bF_{\bg^*\bg^*,t,k}&= \bJ_{\bg_{k,t}}^H(\bR_{t,k}^{-T} \otimes \bR_{t,k}^{-1})\bJ_{\bg^*_{k,t}}\\
	\bF_{\bg\bsigma_r,t,k} & = \bJ_{\bg_{k,t}}^H(\bR_{t,k}^{-T} \otimes \bR_{t,k}^{-1})\bJ_{\bsigma_{r,k,t}}\\
	\bF_{\bg^*\bsigma_r,t,k} & = \bJ_{\bg_{k,t}}^H(\bR_{t,k}^{-T} \otimes \bR_{t,k}^{-1})\bJ_{\bsigma_{r,k,t}}\\
	\bF_{\bsigma_r\bsigma_r,t,k} & = \bJ_{\bsigma_{r,k,t}}^H(\bR_{t,k}^{-T} \otimes \bR_{t,k}^{-1})\bJ_{\bsigma_{r,k,t}}
	\end{align}
	then FIM for the entire data set can be calculated using 
	\begin{equation}
	\bF = \left [\begin{array}{c|c}
	\bF_{cc} & \bF_{cr} \\ \hline
	\bF_{cr}^H & \bF_{rr} 
	\end{array}\right]
	\end{equation}
	where $\bF_{cc} = \sum_{k=1}^K \bE_k$ with
	\[
	\bE_k = \begin{bmatrix}
	\bv_k\bv_k^T \otimes \sum_{t=1}^T \bF_{\bg\bg,t,k} & \bv_k\bv_k^T \otimes \sum_{t=1}^T \bF_{\bg\bg^*,t,k} \\
	\left ( \bv_k\bv_k^T \otimes \sum_{t=1}^T \bF_{\bg\bg^*,t,k} \right)^H & \bv_k\bv_k^T \otimes \sum_{t=1}^T \bF_{\bg^*\bg^*,t,k}
	\end{bmatrix},
	\]
	\[
	\begin{array}{l}
	\bF_{cr}= \begin{bmatrix}
	\bv_1 \otimes \bF_{\bg\bsigma_r,1,1} & \dots & \bv_K \otimes \bF_{\bg\bsigma_r,T,K} \\ 
	\bv_1 \otimes \bF_{\bg^*\bsigma_r,1,1} & \dots & \bv_K \otimes \bF_{\bg^*\bsigma_r,T,K}
	\end{bmatrix},
	\end{array}
	\]
	\[
	\begin{array}{l}
	\bF_{rr} = \bdiag(\bF_{\bsigma_r\bsigma_r,t,k}) \\
	=\begin{bmatrix}
	\bF_{\bsigma_r\bsigma_r,1,1} & & & & \\
	& \ddots && & \\
	&& \bF_{\bsigma_r\bsigma_r,T,1} &&\\
	&&& \ddots & \\
	&&&& \bF_{\bsigma_r\bsigma_r,T,K}
	\end{bmatrix}.
	\end{array}
	\]
	
	For the calculation of the excess noise we need \mbox{$\bF_{cc}-\bF_{cr}\bF_{rr}^{-1}\bF_{cr}^H$} which can be simplified as \eqref{eq:BLOCINVQ}. Computing the CRB with multiple time-snapshot can be parallelized in a great extend given the block-sparse nature of the matrix $\bJ$.
	
\end{document}

%% file: notations.tex
\newcommand{\vect}[0]{\text{vect}}

\newcommand{\vectdiag}[0]{\text{vectdiag}}
\newcommand{\rank}[0]{\text{rank}}

\newcommand{\diag}[0]{\text{diag}}
\newcommand{\bdiag}[0]{\text{bdiag}}

\newcommand{\Cov}[0]{\text{Cov}}

\newcommand{\bGamma}[0]{\boldsymbol{\Gamma}}

\newcommand{\boeta}[0]{\boldsymbol{\eta}}
\newcommand{\bsigmah}[0]{\boldsymbol{\hat{\sigma}}}
\newcommand{\bSigma}[0]{\boldsymbol{\Sigma}}

\newcommand{\ba}[0]{\mathbf{a}}
\newcommand{\bA}[0]{\mathbf{A}}
\newcommand{\bb}[0]{\mathbf{b}}
\newcommand{\bB}[0]{\mathbf{B}}
\newcommand{\bc}[0]{\mathbf{c}}
\newcommand{\bC}[0]{\mathbf{C}}
\newcommand{\bd}[0]{\mathbf{d}}
\newcommand{\bD}[0]{\mathbf{D}}

\newcommand{\bE}[0]{\mathbf{E}}
\newcommand{\bff}[0]{\mathbf{f}}
\newcommand{\bF}[0]{\mathbf{F}}
\newcommand{\bg}[0]{\mathbf{g}}
\newcommand{\bG}[0]{\mathbf{G}}
\newcommand{\bh}[0]{\mathbf{h}}
\newcommand{\bH}[0]{\mathbf{H}}

\newcommand{\bI}[0]{\mathbf{I}}

\newcommand{\bJ}[0]{\mathbf{J}}
\newcommand{\bk}[0]{\mathbf{k}}
\newcommand{\bK}[0]{\mathbf{K}}

\newcommand{\bL}[0]{\mathbf{L}}

\newcommand{\bM}[0]{\mathbf{M}}
\newcommand{\bn}[0]{\mathbf{n}}

\newcommand{\bP}[0]{\mathbf{P}}

\newcommand{\bQ}[0]{\mathbf{Q}}
\newcommand{\br}[0]{\mathbf{r}}
\newcommand{\bR}[0]{\mathbf{R}}
\newcommand{\bs}[0]{\mathbf{s}}
\newcommand{\bS}[0]{\mathbf{S}}

\newcommand{\bU}[0]{\mathbf{U}}
\newcommand{\bv}[0]{\mathbf{v}}
\newcommand{\bV}[0]{\mathbf{V}}
\newcommand{\bw}[0]{\mathbf{w}}
\newcommand{\bW}[0]{\mathbf{W}}
\newcommand{\bx}[0]{\mathbf{x}}
\newcommand{\bX}[0]{\mathbf{X}}
\newcommand{\by}[0]{\mathbf{y}}

\newcommand{\bz}[0]{\mathbf{z}}
\newcommand{\bZ}[0]{\mathbf{Z}}

\newcommand{\EExp}[0]{\mathcal{E}}

\newcommand{\bHt}[0]{\mathbf{\tilde{H}}}
\newcommand{\bJt}[0]{\mathbf{\tilde{J}}}
\newcommand{\bQt}[0]{\mathbf{\tilde{Q}}}
\newcommand{\bRt}[0]{\mathbf{\tilde{R}}}

\newcommand{\bWt}[0]{\mathbf{\tilde{W}}}

\newcommand{\bUt}[0]{\mathbf{\tilde{U}}}
\newcommand{\bVt}[0]{\mathbf{\tilde{V}}}

\newcommand{\bSigmat}[0]{\boldsymbol{\tilde{\Sigma}}}

\newcommand{\bgi}[0]{\boldsymbol{\mathit{g}}}
\newcommand{\brh}[0]{\mathbf{\hat{r}}}
\newcommand{\bPh}[0]{\mathbf{\hat{P}}}

\newcommand{\bRh}[0]{\mathbf{\hat{R}}}

\newcommand{\bthetah}[0]{\boldsymbol{\hat{\theta}}}

\newcommand{\zeros}[0]{\mathbf{0}}
\newcommand{\ones}[0]{\mathbf{1}}

\newcommand{\MCE}[0]{\mathcal{E}}

\newcommand{\MCF}[0]{\mathcal{F}}

\newcommand{\MCN}[0]{\mathcal{N}}

\newcommand{\beq}{\begin{equation}}
\newcommand{\eeq}{\end{equation}}
\newcommand{\bea}{\begin{array}}
\newcommand{\ena}{\end{array}}

\newcommand{\DL}{\begin{dashlist}}
\newcommand{\DLE}{\end{dashlist}}

%% file: CRB_gains.bbl
\begin{thebibliography}{}
\makeatletter
\relax
\def\mn@urlcharsother{\let\do\@makeother \do\$\do\&\do\#\do\^\do\_\do\%\do\~}
\def\mn@doi{\begingroup\mn@urlcharsother \@ifnextchar [ {\mn@doi@}
  {\mn@doi@[]}}
\def\mn@doi@[#1]#2{\def\@tempa{#1}\ifx\@tempa\@empty \href
  {http://dx.doi.org/#2} {doi:#2}\else \href {http://dx.doi.org/#2} {#1}\fi
  \endgroup}
\def\mn@eprint#1#2{\mn@eprint@#1:#2::\@nil}
\def\mn@eprint@arXiv#1{\href {http://arxiv.org/abs/#1} {{\tt arXiv:#1}}}
\def\mn@eprint@dblp#1{\href {http://dblp.uni-trier.de/rec/bibtex/#1.xml}
  {dblp:#1}}
\def\mn@eprint@#1:#2:#3:#4\@nil{\def\@tempa {#1}\def\@tempb {#2}\def\@tempc
  {#3}\ifx \@tempc \@empty \let \@tempc \@tempb \let \@tempb \@tempa \fi \ifx
  \@tempb \@empty \def\@tempb {arXiv}\fi \@ifundefined
  {mn@eprint@\@tempb}{\@tempb:\@tempc}{\expandafter \expandafter \csname
  mn@eprint@\@tempb\endcsname \expandafter{\@tempc}}}

\bibitem[\protect\citeauthoryear{Boonstra \& van~der Veen}{Boonstra \& van~der
  Veen}{2003}]{Boonstra2003}
Boonstra A.-J.,  van~der Veen A.-J.,  2003, IEEE Tr. Signal Processing, 51, 25

\bibitem[\protect\citeauthoryear{Briggs}{Briggs}{1995}]{Briggs1995}
Briggs D.~S.,  1995, PhD thesis, The New Mexico Institute of Mining and
  Technology, Socorro, New Mexico

\bibitem[\protect\citeauthoryear{Brossard, Korso, Pesavento, Boyer, Larzabal
  \& Wijnholds}{Brossard et~al.}{2018}]{Brossard2018}
Brossard M.,  Korso M. N.~E.,  Pesavento M.,  Boyer R.,  Larzabal P.,
  Wijnholds S.~J.,  2018, \mn@doi [Signal Processing]
  {https://doi.org/10.1016/j.sigpro.2017.12.014}, 145, 258

\bibitem[\protect\citeauthoryear{Carrillo, McEwen  \& Wiaux}{Carrillo
  et~al.}{2014}]{Carrillo2014}
Carrillo R.~E.,  McEwen J.~D.,   Wiaux Y.,  2014, Monthly Notices of the Royal
  Astronomical Society, 439, 3591

\bibitem[\protect\citeauthoryear{DeBoer et~al.,}{DeBoer
  et~al.}{2017}]{DeBoer2017}
DeBoer D.~R.,  et~al., 2017, Publications of the Astronomical Society of the
  Pacific, 129, 045001

\bibitem[\protect\citeauthoryear{Dewdney, Hall, Schilizzi  \& Lazio}{Dewdney
  et~al.}{2009}]{Dewdney2009}
Dewdney P.~E.,  Hall P.~J.,  Schilizzi R.~T.,   Lazio T. J.~L.,  2009,
  Proceedings of the IEEE, 97, 1482

\bibitem[\protect\citeauthoryear{Furlanetto, Peng~Oh  \& Briggs}{Furlanetto
  et~al.}{2006}]{Furlanetto2006a}
Furlanetto S.~R.,  Peng~Oh S.,   Briggs F.~H.,  2006, Physics Reports, 433, 181

\bibitem[\protect\citeauthoryear{Hall}{Hall}{2005}]{Hall2005}
Hall P.,  ed. 2005, The Square Kilometer Array: An Engineering Perspective.
Springer

\bibitem[\protect\citeauthoryear{H\"{o}gbom}{H\"{o}gbom}{1974}]{Hoegbom1974}
H\"{o}gbom J.~A.,  1974, Astronomy and Astrophysics Supplement Series, 15, 417

\bibitem[\protect\citeauthoryear{Jagannatham \& Rao}{Jagannatham \&
  Rao}{2004}]{Jagannatham2004}
Jagannatham A.~K.,  Rao B.~D.,  2004, IEEE SIGNAL PROCESSING LETTERS, 11

\bibitem[\protect\citeauthoryear{Kay}{Kay}{1993}]{Kay1993a}
Kay S.~M.,  1993, Fundamentals of Statistical Signal Processing, Estimation
  theory.
 Vol. Volume I, Prentice Hall

\bibitem[\protect\citeauthoryear{Kazemi, Yatawatta  \& Zaroubi}{Kazemi
  et~al.}{2013}]{Kazemi2013}
Kazemi S.,  Yatawatta S.,   Zaroubi S.,  2013, \mn@doi [Monthly Notices of the
  Royal Astronomical Society] {10.1093/mnras/stt1229}, 434, 3130

\bibitem[\protect\citeauthoryear{Koopmans et~al.,}{Koopmans
  et~al.}{2015}]{Koopmans2015}
Koopmans L.,  et~al., 2015, in Proceedings of Advancing Astrophysics with the
  Square Kilometre Array (AASKA14). 9 -13 June. p.~1

\bibitem[\protect\citeauthoryear{Leshem}{Leshem}{2009}]{Leshem2009}
Leshem A.,  2009, Monthly Notices of the Royal Astronomical Society. Submitted.

\bibitem[\protect\citeauthoryear{Leshem \& van~der Veen}{Leshem \& van~der
  Veen}{2000}]{Leshem2000a}
Leshem A.,  van~der Veen A.-J.,  2000, IEEE Trans. on Information Theory,
  Special issue on information theoretic imaging, pp 1730--1747

\bibitem[\protect\citeauthoryear{Mertens, Ghosh  \& Koopmans}{Mertens
  et~al.}{2018}]{Mertens2018}
Mertens F.,  Ghosh A.,   Koopmans L.,  2018, Monthly Notices of the Royal
  Astronomical Society, 478, 3640

\bibitem[\protect\citeauthoryear{Morales \& Wyithe}{Morales \&
  Wyithe}{2010}]{Morales2010}
Morales M.~F.,  Wyithe J. S.~B.,  2010, Annual Review of Astronomy and
  Astrophysics, 48, 127

\bibitem[\protect\citeauthoryear{Mouri~Sardarabadi}{Mouri~Sardarabadi}{2016}]{MouriSardarabadi2016}
Mouri~Sardarabadi A.,  2016, PhD thesis, Delft University of Technology

\bibitem[\protect\citeauthoryear{Mouri~Sardarabadi \& van~der
  Veen}{Mouri~Sardarabadi \& van~der Veen}{2014}]{MouriSardarabadi2014}
Mouri~Sardarabadi A.,  van~der Veen A.-J.,  2014, in 2014 IEEE 8th Sensor Array
  and Multichannel Signal Processing Workshop (SAM). pp 153--156,
  \mn@doi{10.1109/SAM.2014.6882363}

\bibitem[\protect\citeauthoryear{{Mouri Sardarabadi}, {Leshem}  \& {van der
  Veen}}{{Mouri Sardarabadi} et~al.}{2015}]{MouriSardarabadi2015}
{Mouri Sardarabadi} A.,  {Leshem} A.,   {van der Veen} A.-J.,  2015, {Astronomy
  and Astrophysics}

\bibitem[\protect\citeauthoryear{Mouri~Sardarabadi, van~der Veen  \&
  Boonstra}{Mouri~Sardarabadi et~al.}{2016}]{MouriSardarabadi2016a}
Mouri~Sardarabadi A.,  van~der Veen A.-J.,   Boonstra A.-J.,  2016, \mn@doi
  [IEEE Trans. Signal Process.] {10.1109/TSP.2015.2483481}, 64, 432

\bibitem[\protect\citeauthoryear{Mouri~Sardarabadi, van~der Veen  \&
  Koopmans}{Mouri~Sardarabadi et~al.}{2018}]{sardarabadi2018efficient}
Mouri~Sardarabadi A.,  van~der Veen A.-J.,   Koopmans L.~V.,  2018, arXiv
  preprint arXiv:1803.05707

\bibitem[\protect\citeauthoryear{Mouri Sardarabadi \&
  Koopmans}{Mouri Sardarabadi \& Koopmans}{2019}]{MouriSardarabadi2018b}
Mouri Sardarabadi A.,  Koopmans L. V.~E.,  2019, \mn@doi [Monthly Notices of
  the Royal Astronomical Society] {10.1093/mnras/sty3444}, 483, 5480

\bibitem[\protect\citeauthoryear{Offringa et~al.,}{Offringa
  et~al.}{2014}]{Offringa2014}
Offringa A.,  et~al., 2014, \mn@doi [Monthly Notices of the Royal Astronomical
  Society] {10.1093/mnras/stu1368}, 444, 606

\bibitem[\protect\citeauthoryear{{Patil} et~al.,}{{Patil}
  et~al.}{2016}]{Patil2016}
{Patil} A.~H.,  et~al., 2016, \mn@doi [\mnras] {10.1093/mnras/stw2277}, \href
  {http://adsabs.harvard.edu/abs/2016MNRAS.463.4317P} {463, 4317}

\bibitem[\protect\citeauthoryear{Patil et~al.,}{Patil et~al.}{2017}]{Patil2017}
Patil A.,  et~al., 2017, The Astrophysical Journal, 838, 65

\bibitem[\protect\citeauthoryear{Rohde}{Rohde}{1965}]{Rohde1965}
Rohde C.~A.,  1965, Journal of the Society for Industrial and Applied
  Mathematics, 13, 1033

\bibitem[\protect\citeauthoryear{Rothenberg}{Rothenberg}{1971}]{Rothenberg1971}
Rothenberg T.~J.,  1971, Econometrica, 39, pp. 577

\bibitem[\protect\citeauthoryear{Schreier}{Schreier}{2010}]{Schreier2010}
Schreier P.~J.,  2010, Statistical Signal Processing of Complex-Valued Data.
Cambridge University Press

\bibitem[\protect\citeauthoryear{Taguchi}{Taguchi}{1986}]{Taguchi1986}
Taguchi G.,  1986, Introduction to quality engineering: designing quality into
  products and processes

\bibitem[\protect\citeauthoryear{{Trott} \& {Wayth}}{{Trott} \&
  {Wayth}}{2016}]{Trott2016}
{Trott} C.~M.,  {Wayth} R.~B.,  2016, \mn@doi [\pasa] {10.1017/pasa.2016.18},
  \href {http://adsabs.harvard.edu/abs/2016PASA...33...19T} {33, e019}

\bibitem[\protect\citeauthoryear{Vedantham \& Koopmans}{Vedantham \&
  Koopmans}{2016}]{Vedantham2016}
Vedantham H.~K.,  Koopmans L. V.~E.,  2016, \mn@doi [Monthly Notices of the
  Royal Astronomical Society] {10.1093/mnras/stw443}, 458, 3099

\bibitem[\protect\citeauthoryear{Wijnholds}{Wijnholds}{2010}]{Wijnholds2010a}
Wijnholds S.~J.,  2010, PhD thesis, Delft University of Technology

\bibitem[\protect\citeauthoryear{Wijnholds \& van~der Veen}{Wijnholds \&
  van~der Veen}{2008}]{Wijnholds2008}
Wijnholds S.,  van~der Veen A.-J.,  2008, \mn@doi [Selected Topics in Signal
  Processing, IEEE Journal of] {10.1109/JSTSP.2008.2004216}, 2, 613

\bibitem[\protect\citeauthoryear{Wijnholds \& van~der Veen}{Wijnholds \&
  van~der Veen}{2009}]{Wijnholds2009}
Wijnholds S.,  van~der Veen A.-J.,  2009, \mn@doi [Signal Processing, IEEE
  Transactions on] {10.1109/TSP.2009.2022894}, 57, 3512

\bibitem[\protect\citeauthoryear{Yatawatta}{Yatawatta}{2010}]{Yatawatta2010}
Yatawatta S.,  2010, in Sensor Array and Multichannel Signal Processing
  Workshop (SAM), 2010 IEEE. pp 69--72

\bibitem[\protect\citeauthoryear{Yatawatta}{Yatawatta}{2015}]{Yatawatta2015}
Yatawatta S.,  2015, Monthly Notices of the Royal Astronomical Society, 449,
  4506

\bibitem[\protect\citeauthoryear{Yatawatta}{Yatawatta}{2016}]{Yatawatta2016}
Yatawatta S.,  2016, pp 265--269

\bibitem[\protect\citeauthoryear{van Haarlem, Wise, Gunst  et~al.}{van Haarlem
  et~al.}{2013}]{Haarlem2013}
van Haarlem M.~P.,  Wise M.~W.,  Gunst A.~W.,   et~al., 2013, \mn@doi
  [Astronomy \& Applications] {10.1051/0004-6361/201220873}, 556, A2

\bibitem[\protect\citeauthoryear{{van der Tol}}{{van der
  Tol}}{2009}]{vanderTol2009}
{van der Tol} S.,  2009, PhD thesis, Delft University of Technology

\bibitem[\protect\citeauthoryear{van~der Tol, Jeffs  \& van~der Veen}{van~der
  Tol et~al.}{2007}]{Tol2007}
van~der Tol S.,  Jeffs B.,   van~der Veen A.-J.,  2007, \mn@doi [Signal
  Processing, IEEE Transactions on] {10.1109/TSP.2007.896243}, 55, 4497

\makeatother
\end{thebibliography}
